\pdfoutput=1
\documentclass[journal]{IEEEtran}
\usepackage{amsthm,amsmath,amsfonts}
\usepackage{algorithmic}
\usepackage{algorithm}
\usepackage{array}
\usepackage[caption=false,font=normalsize,labelfont=sf,textfont=sf]{subfig}
\usepackage{textcomp}
\usepackage{stfloats}
\usepackage{url}
\usepackage{verbatim}
\usepackage{graphicx}
\usepackage{cite}
\usepackage{multirow}
\usepackage{booktabs}
\usepackage{amssymb}
\begin{document}
\title{Adaptive Weighting Push-SUM for Decentralized Optimization with Statistical Diversity}
\author{Yiming Zhou, Yifei Cheng, Linli Xu, Enhong Chen*,
\IEEEmembership{Fellow, IEEE}
\thanks{* Corresponding author.}
\thanks{All authors are with the Anhui Province Key Laboratory of Big Data Analysis and Application, University of Science and Technology of China (USTC), Hefei, China. E-mail: zym2019@mail.ustc.edu.cn, chengyif@mail.ustc.edu.cn, linlixu@ustc.edu.cn, cheneh@ustc.edu.cn.}
}

\maketitle

\begin{abstract}
Statistical diversity is a property of data distribution and can hinder the optimization of a decentralized network. However, the theoretical limitations of the Push-SUM protocol reduce the performance in handling the statistical diversity of optimization algorithms based on it. In this paper, we theoretically and empirically mitigate the negative impact of statistical diversity on decentralized optimization using the Push-SUM protocol. Specifically, we propose the Adaptive Weighting Push-SUM protocol, a theoretical generalization of the original Push-SUM protocol where the latter is a special case of the former. Our theoretical analysis shows that, with sufficient communication, the upper bound on the consensus distance for the new protocol reduces to $O(1/N)$, whereas it remains at $O(1)$ for the Push-SUM protocol. We adopt SGD and Momentum SGD on the new protocol and prove that the convergence rate of these two algorithms to statistical diversity is $O(N/T)$ on the new protocol, while it is $O(Nd/T)$ on the Push-SUM protocol, where $d$ is the parameter size of the training model. To address statistical diversity in practical applications of the new protocol, we develop the Moreau weighting method for its generalized weight matrix definition. This method, derived from the Moreau envelope, is an approximate optimization of the distance penalty of the Moreau envelope. We verify that the Adaptive Weighting Push-SUM protocol is practically more efficient than the Push-SUM protocol via deep learning experiments.
\end{abstract}


\section{Introduction}
\noindent
\IEEEPARstart{D}{ecentralized} optimization is an important branch of distributed optimization that collaborates nodes in a decentralized network to train a unified model using their local datasets. The collaboration of a decentralized network relies on the specific decentralized communication protocol that can guarantee the synchronization of each node’s parameter or achieve consensus. The Push-SUM protocol \cite{OPS,PerturbedPS,NonConvexPPS}, which can be used in directed decentralized networks, is particularly effective for networks with unstable communication channels. However, as the parameter size of training models rapidly increases, the Push-SUM protocol's ability to handle statistical diversity in decentralized optimization is becoming an issue.

Statistical diversity is a property of data distribution and means that the local datasets on each node are \textbf{not independent and identically distributed} (non-IID). Since the decentralized optimization only communicates nodes' model parameters that are trained on their local dataset, once the local datasets are non-IID, the training across nodes would go in different directions and hinder the convergence of the decentralized network. Statistical diversity is inevitable when the dataset is distributed across each node. Therefore, the negative impact of statistical diversity on optimization is an important issue for Push-SUM-based optimization to address.

Gradient tracking \cite{DIGing,NNLSforGT} is considered an effective solution to address the statistical diversity in the Push-SUM protocol \cite{NonIIDGT}. \texttt{ADD-OPT} \cite{GTSGP} combines Push-SUM with gradient tracking, while \texttt{S-ADDOPT} \cite{SGTSGP} extends full gradient in \texttt{ADD-OPT} to stochastic gradient. However, \texttt{S-ADDOPT} sublinearly converges to the optimum for strongly convex functions, while it is not theoretically effective in addressing the statistical diversity in the Push-SUM protocol when optimizing non-convex functions. On the other hand, introducing the gradient tracking estimators that need to be communicated would tremendously increase the network communication burden. Given that the parameter size of modern DNN models exceeds millions \cite{Resnet}, the communication cost of introduced gradient tracking estimators is becoming unacceptable. Therefore, exploring alternatives to improve the Push-SUM protocol's handling of statistical diversity is valuable.

In this paper, we propose the Adaptive Weighting Push-SUM protocol, which introduced a more flexible definition of the weight matrix that allows the node to set its weights to non-infinitesimal value rather than the reciprocal of outdegree in the original Push-SUM protocol. Our protocol is a generalization of the Push-SUM protocol, and through our theoretical analysis, we reduce the upper bound of the consensus distance from $O(1)$ to $O(1/N)$. We develop \texttt{SGAP} and \texttt{MSGAP} based on our protocol. Our theoretical results show that the upper bound for handling statistical diversity in \texttt{SGAP} is $O(N/T)$, compared to $O(Nd/T)$ for the same algorithm using the original Push-SUM protocol, where $d$ is the parameter size of the training model. For our protocol's applications, we developed a new weighting method, called the Moreau weighting method, from an approximate optimization to the distance penalty of the Moreau envelope \cite{Moreau,Yosida}. We experimentally verify our results, demonstrating that our \texttt{SGAP} and \texttt{MSGAP} are more efficient for statistical diversity in deep learning tasks. Our contributions are summarized as follows.

\begin{itemize}
    \item We propose the Adaptive Weighting Push-SUM protocol, which has a more flexible weighting definition than the original Push-SUM protocol. We improve the theoretical result of the Push-SUM protocol by reducing the upper bound of the consensus distance from $O(1)$ to $O(1/N)$.
    \item We employ SGD and Momentum SGD within our new protocol, thereby developing \texttt{SGAP} and \texttt{MSGAP} respectively, and provide $O\left ( 1/\sqrt{NT} \right )$ convergence rate guarantees for non-convex functions. Specifically, their convergence rate in statistical diversity is $O \left ( N/T \right ) $. In contrast, SGD and Momentum SGD employing the original Push-SUM protocol exhibit a rate of $O \left ( Nd/T \right ) $.
    \item We develop the Moreau weighting method for the flexible definition of the weight matrix in our new protocol. The Moreau weighting method is an approximate optimization of the distance penalty of the Moreau envelope and is more practically efficient in handling statistical diversity.
\end{itemize}

\section{Related Work}
A decentralized optimization algorithm \cite{AcceleratedAB,tcns3,tcns4,PrivacyPS} comprises the decentralized communication protocol and the optimization algorithm. The protocol collaborates the communications in the decentralized network, and the algorithm optimizes the local parameters of each node. Besides the Push-SUM protocol, there are several other protocols \cite{AB,PushPull,ABPushPull,Frost}, but they often make additional assumptions about the network that reduce their robustness in practical applications. However, the theoretical performance of the Push-SUM protocol is not optimal, which limits its application in deep learning, especially for handling statistical diversity \cite{SGP}. Several works \cite{DEXTRA,DIGing,SONATA,Push-LSVRG-UP} have focused on improving the algorithm to address the shortcomings of the Push-SUM-based decentralized optimization algorithm, but they do not tackle the underlying issues of the protocol itself. In this paper, we improve the theoretical performance of the Push-SUM protocol and prove the new protocol is theoretically more efficient in deep learning.

\section{Preliminary}
In this paper, for a vector $\textbf{a}$, $\left\| \textbf{a} \right\|_1$ denotes the L1 norm and $\left\| \textbf{a} \right\|_2$ denotes the L2 norm. For a matrix $\textbf{A}$, $\left\| \textbf{A} \right\|_1$ denotes the L1 norm and $\left\| \textbf{A} \right\|_F$ denotes the Frobenius norm.

\subsection{Network Settings}
We start with defining the key concepts for a directed decentralized network with node scale $N$. The network topology can be described by a sequence of directed graph $\left\{\mathcal{G} ^{(t)}\right\}=\left\{\left[N\right],E^{(t)}\right\} $, where $\left[N\right]=\left\{ 1,2,...,N\right\}$ is the set of nodes and $E^{(t)}\subseteq\left[N\right]\times\left[N\right]$ is the set of communication links. A link $(i,j)\in E^{(t)}$ indicates that node $i$ has a reliable one-way communication channel for sending messages to node $j$ at time $t$. We introduce the following assumption on the directed decentralized network.

\noindent
{\bf Assumption 1} {\it 
Each node in the graph sequence $\left\{\mathcal{G} ^{(t)}\right\}$ has a self-loop. There exist finite positive integers $B$ and $\Delta$, such that the aggregate graph $\bigcup_{i=t}^{t+B-1}\mathcal{G} ^{(i)}$ is strongly connected and each aggregate graph has a diameter at most $\Delta$.
}

\noindent 
Assumption 1 is inherited from the B strongly connected assumption of the original Push-SUM protocol \cite{PerturbedPS,NonConvexPPS}.

Each node has a $N$-dimensional vector that consists of weights for all nodes in the network. These weights are utilized in the consensus step of the protocol for the weighted average computation. Combining every node's weight vector, we get the weight matrix $\textbf{W}^{(t)}=\left [ w_{i,j}^{(t)} \right ] _{N\times N}$. Each weight matrix $\textbf{W}^{(t)}$ is generated according to the graph $\mathcal{G} ^{(t)}$ at time $t$. We introduce the following definitions to the weight matrix.

\noindent
{\bf Definition 1} {\it
$\left\{\textbf{W}^{(t)}\right\}$ is a sequence of weight matrices generated by the graph sequence $\left\{\mathcal{G} ^{(t)}\right\}$ and satisfies the following conditions:
\begin{itemize}
\item[(a)]Each matrix in $\left\{\textbf{W}^{(t)}\right\}$ is a column stochastic matrix and $w^{(t)}_{i,j}>0$ if and only if $\left ( j,i \right ) \in E^{(t)}$, for every $t>0$.
\item[(b)]There exists a scalar $\delta>0$ such that every non-zero scalar $w^{(t)}_{i,j}\ge \delta>0$.
\end{itemize}
}

Compared with the original Push-SUM protocol, Definition 1 allows the entry in $\textbf{W}^{(t)}$ could be any non-infinitesimal value, which is more flexible than the reciprocal of outdegree in the original Push-SUM protocol.

\subsection{Optimization Settings}
We consider the directed decentralized network to solve the following distributed optimization problem:

\noindent
{\bf Problem 1} { \it The algorithm for a decentralized network with $N$ nodes is designed to solve the following minimization problem:
\begin{eqnarray*}\label{Problem}
&& \min_{\textbf{x}\in \mathbb{R}^{d}} \left \{ f\left (\textbf{x}\right ):= \frac{1}{N}\sum_{i=1}^{N}f_{i}\left (\textbf{x}\right ) \right \},\\
&& where\ f_{i}\left (\textbf{x}\right ):= \mathbb{E}_{\xi_{i}\sim D_{i}} \left [ F_{i}\left ( \textbf{x}, \xi_{i}\right ) \right ]. 
\end{eqnarray*}
}

\noindent
In Problem 1, $D_{i}$ is the local data distribution at node $i$, $\xi_{i}$ is a random data sample drawn from $D_{i}$, $F_{i}$ is the local loss function, $f_{i}$ is the expected local loss function over $D_{i}$. 

\noindent
{\bf Definition 2} {\it 
The average parameter of the decentralized network is defined as $\overline{\textbf{x}} = \frac{1}{N} \sum_{i=1}^{N} \textbf{x}_i$.
}

For the expected local loss function $f_{i}$, we introduce the following smoothness assumption.

\noindent
{\bf Assumption 2} {\it 
(\textbf{L-Smooth}): Each function $f_{i}(\textbf{x})$ is \textit{L-smooth} such that
\begin{equation}
\left \| \nabla f_{i}\left ( \textbf{x}_{1}  \right )-\nabla f_{i}\left ( \textbf{x}_{2}  \right )\right \|_{2}\le L\left \| \textbf{x}_{1}-\textbf{x}_{2}  \right \|_{2}, \quad \forall \textbf{x}_{1}, \textbf{x}_{2}\in \mathbb{R}^{d}.
\end{equation}
}

Statistical diversity refers to the phenomenon where the data distributions of local datasets across different nodes are not identical, and is also known as the non-IID. Mathematically, we formalize the difference of data distributions (or the degree of statistical diversity) as the variance of $f_{i}(\textbf{x})$ since it is an expectation of the node's local data distributions. Suppose that variance has an upper bound, we introduce the following assumption to the statistical diversity of Problem 1.

\noindent
{\bf Assumption 3} {\it 
(\textbf{Bounded variances}): There exist $\sigma >0$ and $\kappa >0$ such that for all $\textbf{x}$ and $i$, it follows:
\begin{equation}\label{global_var}
\frac{1}{N}\sum_{i=1}^{N}\left \| \triangledown f_{i}(\textbf{x})-\triangledown f(\textbf{x}) \right \|^{2}_{2}\leq \kappa ^{2},
\end{equation}
\begin{equation}\label{local_var}
\mathbb{E}_{\xi_{i}\sim D_{i}}\left [\left \| \triangledown F_{i}(\textbf{x};\xi_{i} )-\triangledown f_{i}(\textbf{x}) \right \|^{2}_{2} \right ]\leq \sigma ^{2}.
\end{equation}
}

\noindent
The variance $\kappa ^{2}$ is the upper bound of the degree of statistical diversity, and $\sigma ^{2}$ is the upper bound of the sampling variance.


\section{Methodology}

\subsection{Adaptive Weighting Push-SUM Protocol}
\begin{algorithm}[tb]
\caption{Adaptive Weighting Push-SUM Protocol}
\label{AWPS}
\textbf{Input}: The number of iterations $T$. Each node $i$ initializes parameter $\textbf{x}_{i}^{(0)}\in \mathbb{R}^{d}$, corrected parameter $\textbf{y}_{i}^{(0)}=\textbf{x}_{i}^{(0)}$, normalizing scalar $a^{(0)}_{i}=1$, and adjacency parameter buffer $\textbf{x}_{i,j}^{(buff)}=\textbf{x}_{i}^{(0)}$ for $\forall{j} \in \left[N\right]$. \\
\textbf{Output}: $\bar{\textbf{x}}^{(T-1)}$
\begin{algorithmic}[1]
\FOR{$t=\{1, 2, ..., T-1\}$}

\STATE Each node $i$ locally updates its parameter with  the perturbation $\varepsilon ^{(t-1)}_{i}$:

\begin{equation}\label{AWPS1}
\textbf{x}_{i}^{(t-\frac{1}{2})}=\textbf{x}_{i}^{(t-1)}+\varepsilon ^{(t-1)}_{i};
\end{equation}

\STATE Each node $i$ calculates $\left [ w_{1,i}^{(t)}, w_{2,i}^{(t)},...,w_{N,i}^{(t)}  \right ]^{T} $ by the weighting method.
\STATE Each node $i$ pushes $\textbf{x}_{i}^{(t-\frac{1}{2})}$ and $a_{i}^{(t-1)}$ to its neighbors.
\STATE Each node $i$ pushes $w_{j,i}^{(t)}$ to its neighbor $j$.
\STATE Each node $i$ updates adjacency parameter buffer $\textbf{x}_{i,j}^{(buff)}$ for $\forall{j} \in \left[N\right]$:

\begin{equation}\label{bufferUP}
\textbf{x}_{i,j}^{(buff)}=
\left\{
\begin{array}{lr}
\textbf{x}_{j}^{(t-\frac{1}{2} )},\quad\  \mathrm{if}\ (j,i)\in E^{(t)} \\
\textbf{x}_{i}^{(t-\frac{1}{2} )},\quad\  \mathrm{if}\ (j,i)\notin \bigcup_{s=t-B+1}^{t}E^{(s)} \\
\textbf{x}_{i,j}^{(buff)},\quad \mathrm{otherwise}
\end{array}
\right.;
\end{equation}

\STATE Each node $i$ weighted averages the received information

\begin{equation}\label{AWPS2}
\left\{
\begin{array}{lr}
a_{i}^{(t)}=\sum_{j=1}^{N} w_{i,j}^{(t)}a_{j}^{(t-1)}  \\
\textbf{x}_{i}^{(t)}=\sum_{j=1}^{N} w_{i,j}^{(t)}\textbf{x}_{j}^{(t-\frac{1}{2})}
\end{array}
\right.
\quad \forall i,j;
\end{equation}

\STATE Each node $i$ calculates the corrected parameter $\textbf{y}_{i}^{(t)}$

\begin{equation}\label{AWPS3}
\textbf{y}_{i}^{(t)}=\textbf{x}_{i}^{(t)}/a_{i}^{(t)};
\end{equation}

\ENDFOR
\end{algorithmic}
\end{algorithm}

The Adaptive Weighting Push-SUM protocol is described in Algorithm \ref{AWPS}. Lines 3, 4, 5, and 6 are the primal changes compared with the original Push-SUM protocol. These changes are based on the new weight matrix in Definition 1. The highlight of the Adaptive Weighting Push-SUM protocol in Algorithm 1 is it allows the node to assign weights for its neighbors (Line 3) by their previously communicated information, which is maintained in the parameter buffer (Line 6). 

\begin{figure}[htbp]
    \centering
    \includegraphics[width=\linewidth]{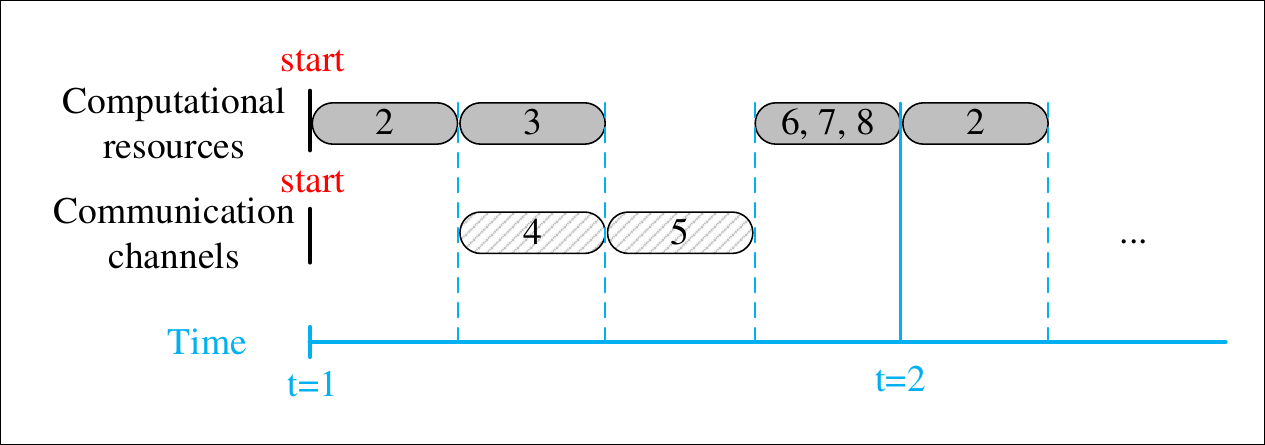}
    \caption{Illustration of Process Arrangement at Node $i$.}
    \label{process_graph}
\end{figure}

In addition, the Adaptive Weighting Push-SUM protocol adopted a novel process arrangement to eliminate the computation delay caused by the computation of the weights. Fig. \ref{process_graph} illustrates this process arrangement, where the number in the figure is the line number in Algorithm 1. Since the computational resources of nodes are free during communication, nodes in Algorithm 1 use their computational resources to calculate the weighting scalars while their communication channels are engaged in transporting parameters and normalizing scalars, followed by the calculated weighting scalars when the communication channels are free. Intuitively, if the nodes' communication bandwidth is limited, the communication time of parameters tends to surpass the weighting time, which fulfills the execution time gap between the new and original protocols. We will discuss the execution time performance of the Adaptive Weighting Push-SUM protocol in experiments.

\subsection{Adaptive Weighting Push-SUM Algorithms}
\begin{algorithm}[tb]
\caption{SGD and Momentum SGD with Adaptive Weighting Push-SUM}
\label{PSMSGD}
\textbf{Input}: The same settings with the Adaptive Weighting Push-SUM protocol, and set learning rate $\gamma >0$, momentum rate $\beta\in [0,1)$, and momentum buffer $\textbf{m}_{i}^{(0)}=\textbf{0}$ for each node $i$.\\
\textbf{Output}: $\bar{\textbf{x}}^{(T-1)}$
\begin{algorithmic}[1] 
\FOR{$t=\{1, 2, ..., T-1\}$}

\STATE Each node $i$ randomly samples $\xi ^{(t-1)}_{i}\sim D_{i}$ and computes $\textbf{g}_{i}^{(t-1)}=\triangledown F_{i}(\textbf{y}^{(t-1)}_{i};\xi^{(t-1)}_{i})$.

\STATE Each node $i$ organized by the Adaptive weighting Push-SUM protocol with the perturbation $\varepsilon ^{(t-1)}_{i}$ as follows:

\begin{flalign}\label{sgap}
& SGAP:\quad \varepsilon ^{(t-1)}_{i}=-\gamma \textbf{g}_{i}^{(t-1)}; &&
\end{flalign}

\begin{flalign}\label{msgd1}
& MSGAP:\quad 
\left\{
\begin{array}{ll}
\textbf{m}_{i}^{(t)}=\beta \textbf{m}_{i}^{(t-1)}+\textbf{g}_{i}^{(t-1)} \\
\varepsilon ^{(t-1)}_{i}=-\gamma \textbf{m}_{i}^{(t)}
\end{array}
\right.;
&&
\end{flalign}\\

\ENDFOR
\end{algorithmic}
\end{algorithm}

The procedure of SGD (\texttt{SGAP}) and Momentum SGD (\texttt{MSGAP}) \cite{Polyak,PolyakFL} with Adaptive Weighting Push-SUM is described in Algorithm 2. Note that when $\beta=0$ in \texttt{MSGAP}, its $\varepsilon ^{(t-1)}_{i}$ becomes identical to that of \texttt{SGAP}, hence we focus solely on the theoretical analysis of \texttt{MSGAP} in this paper.

\texttt{SGAP} and \texttt{MSGAP} are two instantiations of the local update (Line 2 in Algorithm 1) in the Adaptive Weighting Push-SUM protocol. Each node first randomly samples data from its local dataset (Line 2), then computes the gradient term (\texttt{SGAP}) or the momentum term (\texttt{MSGAP}) as the perturbation $\varepsilon ^{(t-1)}_{i}$ (Line 3) in the Adaptive Weighting Push-SUM protocol. Note that the gradient $\textbf{g}_{i}^{(t-1)}$ in Algorithm 2 is calculated based on the corrected parameter $\textbf{y}^{(t-1)}_{i}$, and the output in Algorithm 2 is the average parameter $\bar{\textbf{x}}^{(T-1)}$. After the local update by the perturbation, each node communicates with the network through the remaining steps of the Adaptive Weighting Push-SUM protocol (Line 3 to Line 8 in Algorithm 1).

\subsection{Moreau Weighting Method}
Line 3 in Algorithm 1 is the newly added weighting step of the Adaptive Weighting Push-SUM protocol, which benefits from the flexible definition of the weight matrix in Definition 1. We provide a novel weighting method, named the Moreau weighting method, for the practical application of the new protocol. The Moreau weighting method is the following.

\begin{equation}\label{wPS}
w_{j,i}^{(t)}=\left\{
\begin{aligned}
&1- \frac{2\gamma}{K_{i}^{(t)}}\sum_{(i,s)\in E^{(t)}_{i}} \mathcal{C}'\left ( \left \| \textbf{x}^{(buff)}_{i,i}-\textbf{x}^{(buff)}_{i,s} \right \|^{2}_{2} \right ) \\
&\quad\quad\quad\quad\quad\quad\quad\quad\quad\quad\quad\quad\quad\quad \mathrm{if} \ i=j,  \\
&\frac{2\gamma}{K_{i}^{(t)}} \mathcal{C}'\left ( \left \| \textbf{x}^{(buff)}_{i,i}-\textbf{x}^{(buff)}_{i,j} \right \|^{2}_{2} \right )  \\
&\quad\quad\quad\quad\quad\quad\quad\quad\quad\quad\quad \mathrm{if}\ \left ( i,j \right ) \in E^{(t)}_{i}, \\
&0  \\
&\quad\quad\quad\quad\quad\quad\quad\quad\quad\quad\quad\quad\quad  \mathrm{otherwise}. 
\end{aligned}
\right.
\end{equation}

\noindent where $K_{i}^{(t)}$ is the cardinality of $\left \{ s|(i,s)\in E^{(t)} \right \} $, $E^{(t)}_{i}=E^{(t)}\setminus \{(i,i)\}$. The function $\mathcal{C}'\left ( \cdot  \right ) $ is the following.
\begin{equation}\label{C}
\mathcal{C}'\left ( \left \| \textbf{x}_{1}-\textbf{x}_{2} \right \|^{2}_{2} \right )=\frac{1-v}{2\gamma(1+v)}\left ( 1+v-e^{-k\left \| \textbf{x}_{1}-\textbf{x}_{2} \right \|^{2}_{2}} \right ) ,
\end{equation}

\noindent
where $v$ and $k$ are hyperparameters. Since $\mathcal{C}'\left ( \left \| \textbf{x}_{1}-\textbf{x}_{2} \right \|^{2}_{2} \right ) \in \left [ \frac{1-v}{2\gamma (1+v)} v,\frac{1-v}{2\gamma }  \right )$ ensures $\delta>0$ in Definition 1 exists for given hyperparameter $v$, the weights generated by the Moreau weighting method satisfy Definition 1.

\section{Theoretical Analysis}
\subsection{Consensus Performance of the Protocol}
Following the theoretical analysis of the original Push-SUM protocol \cite{PerturbedPS}, we have the following lemma for the Adaptive Weighting Push-SUM protocol.

\noindent
{\bf Lemma 1} {\it
Suppose that the graph sequence $\{\mathcal{G}^{(t)}\}$ satisfies Assumption 1. Then for the weighting matrix sequence $\{\textbf{W}^{(t)}\}$ defined by Definition 1 and each integer $s\ge 0$, there exists a stochastic vector $\phi(s)$ such that for all $i,j$ and $t\geq s$,
\begin{equation}\label{col_sto_w_equ}
\left | \left [ \textbf{W}^{(t)}\textbf{W}^{(t-1)}...\textbf{W}^{(s+1)}\textbf{W}^{(s)} \right ]_{i,j}-\phi_{i}(t) \right |\leq k\lambda ^{t-s},
\end{equation}
where
\begin{equation}\label{col_sto_w_num}
k=2,\quad \lambda=(1-\delta^{\Delta B})^{\frac{1}{\Delta B}}.
\end{equation}
}

\noindent
Lemma 1 is similar to the theoretical result of the original Push-SUM protocol \cite{PerturbedPS}, indicating that the product of a sequence of column-stochastic matrices converges to a matrix with identical column vectors.

The key theoretical improvement of this paper is the new row sum of the product of a sequence of column-stochastic matrices. We find this row sum has different lower bounds for different sequence lengths. For the sequence with a sufficiently large length, its row sum has a lower bound related to the network scale $N$, which is critical to the theoretical improvement of the Adaptive Weighting Push-SUM protocol. We state and prove this property in the following lemma.

\noindent
{\bf Lemma 2} {\it
Suppose that the graph sequence $\{\mathcal{G}^{(t)}\}$ satisfies Assumption 1. Then for the weighting matrix sequence $\{\textbf{W}^{(t)}\}$ defined by Definition 1, for every $t < \Delta B$
\begin{equation}\label{wm_row_sum_min}
\min_{i} \left [ \textbf{W}^{(t)}\textbf{W}^{(t-1)}...\textbf{W}^{(1)}\textbf{1}  \right ]_{i}\ge \delta^{\Delta B},
\end{equation}
for every $t \ge \Delta B$
\begin{equation}\label{wm_row_sum_max}
\min_{i} \left [ \textbf{W}^{(t)}\textbf{W}^{(t-1)}...\textbf{W}^{(1)}\textbf{1}  \right ]_{i}\ge N \delta^{\Delta B},
\end{equation}
where $\textbf{1}$ is a vector with every entry equals to 1.
}

\begin{proof}
Let $\textbf{W}[t:s]$ denote $\textbf{W}^{(t)}\textbf{W}^{(t-1)}...\textbf{W}^{(s)}$. By Assumption 3 that each node always has a self-loop and the Definition 1 to matrices $\left\{\textbf{W}^{(t)}\right\}$, for every $t<\Delta B$ we have
\begin{equation}\label{lemma3/1}
\left [ \textbf{W}[t:1] \textbf{1}\right ]_{i} \ge \textbf{W}[t:1]_{ii}\ge  \delta^{t}\ge \delta^{\Delta B}.
\end{equation}

For every $t \ge \Delta B$, considering Assumption 3, every aggregate graph with length $B$ is strongly connected and with a diameter at most $\Delta$. Thus, for $t \ge \Delta B$, every entry of $\textbf{W}[t:t-\Delta B +1]$ is positive and has a value at least $\delta^{\Delta B}$. Then, for every entry of $\textbf{W}[t:1]$, we have
\begin{equation}\label{lemma3/2}
\begin{aligned}
&&\left [ \textbf{W}[t:1] \right ]_{ij}&=\left [ \textbf{W}[t:t-\Delta B +1]\textbf{W}[t-\Delta B:1] \right ]_{ij}\nonumber \\
&&\ &=\sum_{k}\textbf{W}[t:t-\Delta B +1]_{ik}\textbf{W}[t-\Delta B:1]_{kj}.
\end{aligned}
\end{equation}

\noindent
where $\textbf{W}[t-\Delta B:1]$ is a column stochastic matrix, thus Eq.(\ref{lemma3/2}) is a convex combination to the $i$'th row of $\textbf{W}[t:t-\Delta B +1]$. Since a convex combination doesn't make the vector's entry smaller, we have

\begin{equation}\label{lemma3/3}
\min_{i,j}\left [ \textbf{W}[t:1] \right ]_{ij} \ge \min_{i,j} \textbf{W}[t:t-\Delta B +1]_{ij}\ge \delta^{\Delta B}.
\end{equation}

\noindent
Then, for every $t \ge \Delta B$, we have

\begin{equation}\label{lemma3/4}
\left [ \textbf{W}[t:1] \textbf{1}\right ]_{i} \ge N \min_{i,j}\left [ \textbf{W}[t:1] \right ]_{ij}\ge N \delta^{\Delta B}.
\end{equation}

\end{proof}

Now we provide the theoretical analysis to the consensus performance of the Adaptive Weighting Push-SUM protocol. We start with a succinct formulation for the protocol and define the following matrices and vector.

\noindent
{\bf Definition 3} {\it
Combining parameters $\textbf{x}_{i}^{(t)}$, $\varepsilon_{i} ^{(t)}$, $\textbf{y}_{i}^{(t)}$, and $a_{i} ^{(t)}$ of all nodes $i\in [N]$ at time $t$, we have the following matrices and vector for the network.
\begin{equation}\label{mp_vector_matrix}
\begin{array}{lr}
\textbf{X}^{(t)}=\left[\textbf{x}_{1}^{(t)},\textbf{x}_{2}^{(t)},...,\textbf{x}_{N}^{(t)}\right]^{T},\ 
\varepsilon^{(t)}=\left[\varepsilon_{1} ^{(t)},\varepsilon_{2} ^{(t)},...,\varepsilon_{N} ^{(t)}\right]^{T},\\
\textbf{Y}^{(t)}=\left[\textbf{y}_{1}^{(t)},\textbf{y}_{2}^{(t)},...,\textbf{y}_{N}^{(t)}\right]^{T},\ 
\textbf{a}^{(t)}=\left[a_{1} ^{(t)},a_{2} ^{(t)},...,a_{N} ^{(t)}\right]^{T},
\end{array}
\end{equation}}

\noindent
where $\textbf{a}^{(t)}$ is a $N$-dimensional vector, and others are the $N\times d$-dimensional matrix. Executing the Adaptive Weighting Push-SUM protocol at time $t$ is equivalent to running the following equations step by step:
\begin{equation}\label{mp_matrix}
\left\{
\begin{array}{lr}
\textbf{X}^{(t-\frac{1}{2})}=\textbf{X}^{(t-1)}+\varepsilon ^{(t-1)}\\
\textbf{a}^{(t)}=\textbf{W}^{(t)}\textbf{a}^{(t-1)}\\
\textbf{X}^{(t)}=\textbf{W}^{(t)}\textbf{X}^{(t-\frac{1}{2})}\\
\left [ \textbf{Y}^{(t)} \right ]_{i}= \left [ \textbf{X}^{(t)} \right ]_{i}/a_{i}^{(t)}
\end{array}
\right.,
\end{equation}

\noindent
where $\left [ \textbf{X} \right ]_{i}$ is the $i$-th row vector of $\textbf{X}_{i}$.

Leveraging the theoretical improvement in Lemma 2, we establish the following upper bound on the distance between a node's corrected parameter and the network average parameter. This upper bound serves as the measure of consensus performance for the Adaptive Weighting Push-SUM protocol.

\noindent
{\bf Theorem 1} {\it
Suppose that the graph sequence $\{\mathcal{G}^{(t)}\}$ satisfies Assumption 1, and the weighting matrix sequence $\{\textbf{W}^{(t)}\}$ is defined by Definition 1. For $t<\Delta B$, the distance between the node's corrected parameter and the network average parameter is bounded by:
\begin{eqnarray}\label{the1_equ}
&&\left \| \textbf{y} _{i}^{(t)}-\frac{1}{N}\sum_{i=1}^{N}\textbf{x}^{(t)}_{i}\right \|_{1} \nonumber \\
&&\le C\lambda^{t-1}\left \|\textbf{X}^{(0)} \right \|_{1}+C\sum_{s=1}^{t}\lambda^{t-s}\left \|\varepsilon^{(s)}\right \|_{1}.\nonumber
\end{eqnarray}

\noindent
For $t\ge \Delta B$:
\begin{eqnarray}\label{the1_equ2}
&&\left \| \textbf{y} _{i}^{(t)}-\frac{1}{N}\sum_{i=1}^{N}\textbf{x}^{(t)}_{i}\right \|_{1} \nonumber \\
&&\le \frac{C}{N}\lambda^{t-1}\left \|\textbf{X}^{(0)} \right \|_{1}+\frac{C}{N}\sum_{s=1}^{t}\lambda^{t-s}\left \|\boldsymbol{\varepsilon}^{(s)}\right \|_{1},\nonumber
\end{eqnarray}
where $C=\frac{4}{\delta^{\Delta B}}$, $\lambda=(1-\delta^{\Delta B})^{\frac{1}{\Delta B}}\in (0,1)$.
}

\begin{proof}
Let $\textbf{W}[t:s]$ denote $\textbf{W}^{(t)}\textbf{W}^{(t-1)}...\textbf{W}^{(s)}$. From Eq.(\ref{mp_matrix}), we have
\begin{eqnarray}\label{glo_loc_proof_1}
\textbf{X}^{(t)}&=&\textbf{W}^{(t)}\textbf{X}^{(t-1)}+\textbf{W}^{(t)}\boldsymbol{\varepsilon}^{(t)} \nonumber \\
~&=&\textbf{W}[t:1]\textbf{X}^{(0)}+\sum_{s=1}^{t}\textbf{W}[t:s] \boldsymbol{\varepsilon} ^{(s)}.
\end{eqnarray}
\noindent
Multiply both sides of Eq.\eqref{glo_loc_proof_1} by $\boldsymbol{\phi}(t)\left (\textbf{1}\right )^{T}$, where $\boldsymbol{\phi} (t)$ is the vector that satisfies Lemma 1, and $\textbf{1}$ is a vector with every entry equal to 1.
\begin{eqnarray}\label{glo_loc_proof_2}
&&\boldsymbol{\phi} (t)\left (\textbf{1}\right )^{T}\textbf{X}^{(t)}\nonumber \\
&&=\boldsymbol{\phi} (t)\left (\textbf{1}\right )^{T}\textbf{W}[t:1]\textbf{X}^{(0)}+\sum_{s=1}^{t}\boldsymbol{\phi} (t)\left (\textbf{1}\right )^{T}\textbf{W}[t:s] \boldsymbol{\varepsilon} ^{(s)} \nonumber \\
&&\overset{(a)}{=} \boldsymbol{\phi} (t)\left (\textbf{1}\right )^{T}\textbf{X}^{(0)}+\sum_{s=1}^{t}\boldsymbol{\phi} (t)\left (\textbf{1}\right )^{T} \boldsymbol{\varepsilon} ^{(s)}.
\end{eqnarray}
(a): $\textbf{W}[t:s]$ is a column stochastic matrix.

\noindent
Combine Eq.\eqref{glo_loc_proof_1} and Eq.\eqref{glo_loc_proof_2} yields that
\begin{eqnarray}\label{glo_loc_proof_3}
&&\textbf{X}^{(t)}-\boldsymbol{\phi} (t)\left (\textbf{1}\right )^{T}\textbf{X}^{(t)}\nonumber \\
&&=\textbf{W}[t:1]\textbf{X}^{(0)}-\boldsymbol{\phi} (t)\left (\textbf{1}\right )^{T}\textbf{X}^{(0)} \nonumber \\
&&\quad +\sum_{s=1}^{t}\textbf{W}[t:s] \boldsymbol{\varepsilon} ^{(s)} -\sum_{s=1}^{t}\boldsymbol{\phi} (t)\left(\textbf{1}\right )^{T} \boldsymbol{\varepsilon} ^{(s)}\nonumber \\
&&\overset{(a)}{=}\textbf{D}[t:1]\textbf{X}^{(0)}+\sum_{s=1}^{t}\textbf{D}[t:s]\boldsymbol{ \varepsilon}^{(s)},
\end{eqnarray}
where $\textbf{D}[t:s] := \textbf{W}[t:s] - \boldsymbol{\phi}(t)\left (\textbf{1}\right )^{T}$.

\noindent
Rearrange Eq.\eqref{glo_loc_proof_3}, we have
\begin{equation}\label{glo_loc_proof_4}
\textbf{X}^{(t)}=\boldsymbol{ \phi }(t)\left (\textbf{1}\right )^{T}\textbf{X}^{(t)}+\textbf{D}[t:1]\textbf{X}^{(0)}+\sum_{s=1}^{t}\textbf{D}[t:s]\boldsymbol{ \varepsilon}^{(s)}.
\end{equation}

\noindent
In the same way, we have
\begin{eqnarray}\label{glo_loc_proof_5}
\textbf{a}^{(t)}&=&\boldsymbol {\phi} (t)\left (\textbf{1}\right )^{T}\textbf{W}[t:1]\textbf{a}^{(0)}+\textbf{D}[t:1]\textbf{a}^{(0)}\nonumber \\
&\overset{(a)}{=}&\boldsymbol {\phi} (t)N+\textbf{D}[t:1]\textbf{1}.
\end{eqnarray}
(a): $\textbf{W}[t:1]$ is a column stochastic matrix and $\textbf{a}^{(0)}=\textbf{1}$.

\noindent
Substitute Eq.\eqref{glo_loc_proof_4} and Eq.\eqref{glo_loc_proof_5} into $\left ( \textbf{y}_{i}^{(t)} \right ) ^T=\left [ \textbf{Y} \right ]_{i}^{(t)}=\left [ \textbf{X} \right ]^{(t)}_{i}/a^{(t)}_{i}$ in Eq.\eqref{mp_matrix} yields that
\begin{equation}\label{glo_loc_proof_6}
\begin{aligned}
&\left ( \textbf{y}_{i}^{(t)} \right ) ^T\nonumber\\
&=\frac{\left [\boldsymbol{\phi} (t)\left (\textbf{1}\right )^{T}\right ]_{i}\textbf{X}^{(t)}+\left [ \textbf{D}[t:1] \right ]_{i}\textbf{X}^{(0)}+\sum_{s=1}^{t}\left [ \textbf{D}[t:s] \right ]_{i}\boldsymbol{\varepsilon}^{(s)}}{\boldsymbol{\phi}_{i} (t)N+\left [ \textbf{D}[t:1]\textbf{1}\right ]_{i}}.
\end{aligned}
\end{equation}

\noindent
For the gap between $\left ( \textbf{y}_{i}^{(t)} \right ) ^T$ and $\frac{\textbf{1}^{T} }{N}X^{(t)}$, we have
\begin{equation}\label{glo_loc_proof_7}
\begin{aligned}
&\left ( \textbf{y}_{i}^{(t)} \right ) ^T-\frac{\textbf{1}^{T} }{N}\textbf{X}^{(t)}\nonumber\\
&=\frac{\left [\boldsymbol{\phi} (t)\left (\textbf{1}\right )^{T}\right ]_{i}\textbf{X}^{(t)}+\left [ \textbf{D}[t:1] \right ]_{i}\textbf{X}^{(0)}+\sum_{s=1}^{t}\left [ \textbf{D}[t:s] \right ]_{i}\boldsymbol{\varepsilon}^{(s)}}{\boldsymbol{\phi}_{i} (t)N+\left [ \textbf{D}[t:1]\textbf{1}\right ]_{i}}\nonumber \\
&\quad -\frac{\textbf{1}^{T} }{N}\textbf{X}^{(t)}\nonumber \\
&\overset{(a)}{=}\frac{N\left [ \textbf{D}[t:1] \right ]_{i}\textbf{X}^{(0)}+N\sum_{s=1}^{t}\left [ \textbf{D}[t:s] \right ]_{i}\boldsymbol{\varepsilon}^{(s)}}{N\left (\boldsymbol{\phi}_{i} (t)N+\left [ \textbf{D}[t:1]\textbf{1}\right ]_{i}\right )}\nonumber \\
&\quad -\frac{\textbf{1}^{T}\textbf{X}^{(t)}\left [ \textbf{D}[t:1]\textbf{1}\right ]_{i}}{N\left (\boldsymbol{\phi}_{i} (t)N+\left [ \textbf{D}[t:1]\textbf{1}\right ]_{i}\right )}.
\end{aligned}
\end{equation}
(a): $\left [ \boldsymbol{\phi} (t)\left (\textbf{1}\right )^{T}\right ]_{i}=\boldsymbol{\phi}_{i}(t)\left(\textbf{1}\right)^{T}$.

\noindent
Take the L1 vector norm to both sides of Eq.\eqref{glo_loc_proof_7}. For $t\ge \Delta B$,

\begin{equation}\label{glo_loc_proof_8}
\begin{aligned}
&\left \| \left ( \textbf{y}_{i}^{(t)} \right ) ^T-\frac{\textbf{1}^{T} }{N}\textbf{X}^{(t)} \right \| _{1} \\
&\overset{(a)}{\le }\frac{\left \| \left [ \textbf{D}[t:1] \right ]_{i}\textbf{X}^{(0)} \right \|_{1}}{N\delta ^{\Delta B}}+\frac{1}{N\delta ^{\Delta B}}\sum_{s=1}^{t}\left \| \left [ \textbf{D}[t:s] \right ]_{i}\boldsymbol{\varepsilon}^{(s)}\right \|_{1} \\
&\quad +\frac{1}{N^{2}\delta ^{\Delta B}}\left \| \textbf{1}^{T}\textbf{X}^{(t)}\left [ \textbf{D}[t:1]\right ]_{i}\textbf{1} \right \|_{1}\\
&\overset{(b)}{\le }\frac{1}{N\delta ^{\Delta B}} \max\left \{ |\left [ \textbf{D}[t:1] \right ]_{i} |\right \}  \left \|\textbf{X}^{(0)} \right \|_{1}\\
&\quad +\frac{1}{N\delta ^{\Delta B}}\sum_{s=1}^{t}\max\left \{ |\left [ \textbf{D}[t:s] \right ]_{i} |\right \}\left \|\boldsymbol{\varepsilon}^{(s)}\right \|_{1} \\
&\quad +\frac{1}{N\delta ^{\Delta B}}\max\left \{ |\left [ \textbf{D}[t:1] \right ]_{i} |\right \} \left \| \textbf{1}^{T}\textbf{X}^{(t)} \right \|_{1}\\
&\overset{(c)}{\le }\frac{k\lambda^{t-1}}{N\delta ^{\Delta B}}   \left \|\textbf{X}^{(0)} \right \|_{1}+\frac{k}{N\delta ^{\Delta B}}\sum_{s=1}^{t}\lambda^{t-s}\left \|\boldsymbol{\varepsilon}^{(s)}\right \|_{1} \\
&\quad +\frac{1}{N\delta ^{\Delta B}}k\lambda^{t-1}\left \| \textbf{1}^{T}\textbf{X}^{(t)} \right \|_{1}.
\end{aligned}
\end{equation}
(a): By $\boldsymbol{\phi}_{i}(t)N+\left [ \textbf{D}[t:1]\textbf{1}\right ]_{i}=\left [ \textbf{W}[t:1]\textbf{1}\right ]_{i}$ and Lemma 2 with $t \geq \Delta B$.\par
\noindent
(b): By $\left \| \textbf{a}\textbf{A}  \right \| _{1}\le \max \{\left | \textbf{a} \right | \} \left \|\textbf{A}  \right \| _{1}$, where $\max \{\left | \textbf{a} \right | \} $ is the entry of vector $\textbf{a}$ with the largest absolute value.\par
\noindent
(c): By Lemma 1.

\noindent
For $\left \| \textbf{1}^{T}\textbf{X}^{(t)} \right \|_{1}$ in Eq.\eqref{glo_loc_proof_8},
\begin{eqnarray}\label{glo_loc_proof_9}
\left \| \textbf{1}^{T}\textbf{X}^{(t)} \right \|_{1}&\overset{(a)}{=}&\left \| \textbf{1}^{T}\textbf{X}^{(0)}+\sum_{s=1}^{t}\textbf{1}^{T}\boldsymbol{\varepsilon}^{(s)} \right \|_{1} \nonumber \\
&\overset{(b)}{\le }&\left \| \textbf{X}^{(0)} \right \|_{1}+\sum_{s=1}^{t}\left \| \boldsymbol{\varepsilon}^{(s)} \right \|_{1}.
\end{eqnarray}
(a): $\textbf{W}[t:s]$ is a column stochastic matrix.\par
\noindent
(b): By $\left \| \textbf{a}\textbf{A} \right \|_{1} \leq \max \{\left | \textbf{a} \right | \} \left \|\textbf{A} \right \|_{1}$.

\noindent
Take Eq.\eqref{glo_loc_proof_9} into Eq.\eqref{glo_loc_proof_8}, we have
\begin{eqnarray}\label{glo_loc_proof_10}
&&\left \| \left ( \textbf{y}_{i}^{(t)} \right ) ^T-\frac{\textbf{1}^{T} }{N}\textbf{X}^{(t)} \right \| _{1}\nonumber \\
&&\overset{(a)}{\le} \frac{2k\lambda^{t-1}}{N\delta ^{\Delta B}}\left \|\textbf{X}^{(0)} \right \|_{1}+\frac{2k}{N\delta ^{\Delta B}}\sum_{s=1}^{t}\lambda^{t-s}\left \|\boldsymbol{\varepsilon}^{(s)}\right \|_{1}\nonumber\\
&&\overset{(b)}{\le} \frac{C}{N}\lambda^{t-1}\left \|\textbf{X}^{(0)} \right \|_{1}+\frac{C}{N}\sum_{s=1}^{t}\lambda^{t-s}\left \|\boldsymbol{\varepsilon}^{(s)}\right \|_{1}.
\end{eqnarray}
(a): $\lambda<1$.\par
\noindent
(b): Let $C:=\frac{2k}{\delta^{\Delta B}}=\frac{4}{\delta^{\Delta B}}$.\par

\noindent
For $t< \Delta B$, take Lemma 2 into Eq.(\ref{glo_loc_proof_8}), and in the same as $t\ge \Delta B$, we can prove Theorem 1 for $t< \Delta B$.
\end{proof}\bigskip

\noindent
\textbf{Comparison with the Original Push-SUM Protocol}

Following the same settings as in the original Push-SUM protocol \cite{PerturbedPS}, where the minimum entry is set to $\delta = \frac{1}{N}$ and the graph diameter to $\Delta = N$, we derive the values $C = 4N^{NB}$ and $\lambda=\left ( 1-\frac{1}{N^{NB}} \right ) ^{\frac{1}{NB}} $ in Theorem 1. These values are nearly the same as the multiplier $8N^{NB}$ and the base $\left ( 1-\frac{1}{N^{NB}} \right ) ^{\frac{1}{NB}}$ in the original Push-SUM protocol (Lemma 1 in \cite{PerturbedPS}). For $t<\Delta B$, comparing the theoretical results of the new and original protocols under the same settings, they are nearly the same result.  For $t\ge \Delta B$, comparing the theoretical results of the new and original protocols under the same settings, our upper bound in Theorem 1 is $O\left ( \frac{1}{N}  \right ) $, while the upper bound of the original protocol remains the same as for $t<\Delta B$ and is $O\left ( 1 \right )$. Considering the network scale $N$ is far beyond $1$, the consensus performance of the Adaptive Weighting Push-SUM protocol is better than the original Push-SUM protocol.


\subsection{Optimization Performance of the Algorithms}

Based on Theorem 1, we provide the following theoretical convergence guarantee for our \texttt{MSGAP} (Algorithm 2).

\noindent
{\bf Theorem 2} {\it 
Suppose that Assumption 1, 2, 3 hold, under Definition 1, if the step-size
$\gamma\leq min\left \{ \frac{(1-\alpha)^{2}}{12\sqrt{2}CNL},\frac{(1-\beta)^{2}}{2L(1+\beta)} \right \}$, where $\alpha\in (0,1)$, for \texttt{MSGAP} and all $T>\Delta B$, we have
\begin{equation}\label{theorem3eql}
\begin{aligned}
&\frac{1}{T}\sum_{t=0}^{T-1}\mathbb{E}\left [ \left \| \triangledown f(\overline{\textbf{x}}^{(t)}) \right \|^{2}_{2} \right ]\nonumber \\
&\leq \frac{2-2\beta}{\gamma T}\left ( f(\overline{\textbf{x}}^{(0)} )- f^{*}\right )+\left(\frac{4L^{2}}{T}+\frac{12C^{2}NL^{2}}{T(1-\lambda^{2})}\right)\left \| \textbf{X}^{(0)}\right \|^{2}_{F}\\
&\quad+\frac{\gamma L}{N(1-\beta)^{2}}\sigma^{2}+\frac{18\gamma^{2}C^{2}L^{2}}{(1-\alpha)^{4}}\left ( 1+\frac{N^{2}\Delta B}{T} \right ) \sigma^{2}\\
&\quad +\frac{72\gamma^{2}C^{2}L^{2}}{(1-\alpha)^{4}}\left ( 1+\frac{N^{2}\Delta B}{T} \right ) \kappa ^{2}.
\end{aligned}
\end{equation}
}

\noindent
The proof of Theorem 2 is in the Appendix. With an appropriately chosen step-size $\gamma$, we can get the following corollary.

\noindent
{\bf Corollary 1} {\it 
Under the same conditions as in Theorem 2, for \texttt{MSGAP} with sufficiently large $T$ and $\gamma=O\left ( \sqrt{\frac{N}{T}}\right )$, we have
\begin{eqnarray*}
&&\frac{1}{T}\sum_{t=0}^{T-1}\mathbb{E}\left [ \left \| \triangledown f(\overline{\textbf{x}}^{(t)}) \right \|^{2}_{2} \right ]\\
&&\leq O\left(\frac{1}{\sqrt{NT}}\right) +O\left ( \frac{\sigma^{2}}{\sqrt{NT}} \right )+O\left ( \frac{N\kappa^{2}}{T} \right ).
\end{eqnarray*}
}

\noindent
This result indicates that with \texttt{MSGAP}, the average parameter $\overline{\textbf{x}}^{(t)}$ will converge to the solution of Problem 1. And for fixed $N$, the $O\left(\frac{1}{\sqrt{NT}}\right)$ term will dominate the other factors in Corollary 1 when $T$ is sufficiently large. Therefore, for a decentralized network satisfying Assumption 1, our \texttt{MSGAP} algorithm guarantees it will converge to the solution of Problem 1 at a speed of $O\left(\frac{1}{\sqrt{NT}}\right)$ after sufficient iterations.

\noindent
{\bf Corollary 2} {\it 
Under the same assumptions as in Theorem 2, for \texttt{SGAP} with sufficiently large $T$, if the step-size $\gamma\leq min\left \{ \frac{(1-\lambda)^{2}}{12\sqrt{2}CNL},\frac{1}{2L} \right \}$ and $\gamma=O\left ( \sqrt{\frac{N}{T}}\right )$, we have
\begin{equation*}
\begin{aligned}
&\frac{1}{T}\sum_{t=0}^{T-1}\mathbb{E}\left [ \left \| \triangledown f(\overline{\textbf{x}}^{(t)}) \right \|^{2}_{2} \right ]\\
&\leq \frac{2}{\gamma T}\left ( f(\overline{\textbf{x}}^{(0)} )- f^{*}\right )+ \left(\frac{4L^{2}}{T}+\frac{12C^{2}NL^{2}}{T(1-\lambda^{2})}\right)\left \| \textbf{X}^{(0)}\right \|^{2}_{F}\\
&\quad+\frac{\gamma L}{N}\sigma^{2}+\frac{18\gamma^{2}C^{2}L^{2}}{(1-\lambda)^{4}}\left ( 1+\frac{N^{2}\Delta B}{T} \right ) \sigma^{2}\\
&\quad +\frac{72\gamma^{2}C^{2}L^{2}}{(1-\lambda)^{4}}\left ( 1+\frac{N^{2}\Delta B}{T} \right ) \kappa ^{2}\\
&\leq O\left(\frac{1}{\sqrt{NT}}\right) +O\left ( \frac{\sigma^{2}}{\sqrt{NT}} \right )+O\left ( \frac{N\kappa^{2}}{T} \right ).
\end{aligned}
\end{equation*}
}

\noindent
Let $\beta=0$ in Theorem 2, we get Corollary 2.\bigskip

\noindent
\textbf{Comparison with Algorithm Based on the Original Push-SUM Protocol}

\texttt{SGP} and \texttt{SGAP} share the same conditions, and both use SGD as their optimization algorithm. Their only difference is in the protocol used: \texttt{SGP} employs the original Push-SUM protocol, while \texttt{SGAP} utilizes our Adaptive Weighting Push-SUM protocol. Therefore, we compare \texttt{SGP} and \texttt{SGAP} to show that our protocol is theoretically superior to the original protocol in handling statistical diversity in decentralized optimization.

\noindent
{\bf Corollary 3} {\it 
Suppose that Assumptions 1, 2, and 3 all hold, for \texttt{SGP} \cite{SGP} with sufficiently large $T$, if the step-size is sufficiently small and $\gamma=O\left ( \sqrt{\frac{N}{T}}\right )$, we have

\begin{equation*}
\begin{aligned}
&\frac{1}{T}\sum_{t=1}^{T-1}\mathbb{E}\left [ \left \| \triangledown f(\overline{\textbf{x}}^{(t)}) \right \|^{2}_{2} \right ]\\
&\leq O\left(\frac{1}{\sqrt{NT}}\right) +O\left ( \frac{\sigma^{2}}{\sqrt{NT}} \right )+O\left ( \frac{NP^{2}\kappa^{2}}{T} \right ),
\end{aligned}
\end{equation*}

\noindent
where $P<\frac{2\sqrt{d}K_{max}^{\Delta B}}{\left ( 1-N K_{max}^{-\Delta B}\right ) ^{\frac{\Delta B+2}{\Delta B+1}}}$, and $K_{max}=\max_{i,t}\left \{ K_{i}^{(t)} \right \} $ is the maximum out-degree in the graph sequence $\left\{\mathcal{G} ^{(t)}\right\}$.
}

\noindent
Corollary 3 is the performance of \texttt{SGP} under our symbols. 

First, the upper bound of $P$ in \texttt{SGP} does not exist for networks that satisfy Assumption 1 and meet the assumptions of \texttt{SGP}. Considering a fully connected network, its strongly connected integer $B=1$, the graph diameter integer $\Delta=1$, and $K_{max}=N$. Then for the denominator of the upper bound of $P$, we have $1-N K_{max}^{-\Delta B}=0$. This implies that the upper bound of $P$ does not exist, leading to the theoretical divergence of \texttt{SGP} in a fully connected network.

Second, considering the statistical diversity term $\kappa^2$ in \texttt{SGP}, since the upper bound of $P$ is $O(\sqrt{d})$, the term $O\left(\frac{NP^2\kappa^2}{T}\right)$ simplifies to $O\left(\frac{Nd\kappa^2}{T}\right)$, which is dependent on the parameter size $d$ of the training model. Meanwhile, the statistical diversity term $\kappa^2$ in \texttt{SGAP} is $O\left(\frac{N\kappa^2}{T}\right)$, which mitigates the influence of $d$. Since \texttt{SGP} and \texttt{SGAP} differ only in the protocol, our protocol is theoretically more efficient in handling statistical diversity in decentralized optimization, particularly for the optimization of DNN models with millions of parameters.

\subsection{Derivation of Moreau Weighting}
Our theoretical results show that our protocol is theoretically more efficient than the original protocol in handling statistical diversity in decentralized optimization. However, our improvements are focused on theory, which does not quite work in practice. Fortunately, our protocol has more flexible weight definitions in Definition 1, which are not allowed in the original protocol, enabling us to enhance our protocol by the weighting method. Here, we provide our Moreau weighting method derivative from the Moreau envelope.

The Moreau envelope is a smoothing technique that can turn any convex lower semicontinuous function into a smooth function and an ill-conditioned smooth convex function into a well-conditioned smooth convex function \cite{MoreauSmooth}. Early works \cite{mePFL,Ditto} empirically show that replacing the loss function with its Moreau envelope is effective for mitigating statistical diversity in a centralized network. We find that the optimization of the distance penalty of the Moreau envelope at each node can be approximated as a convex combination of its neighbors' parameters, which can be simplified to a weighting method. We start with the analysis of the Moreau envelope.

\noindent
{\bf Remark 1} {\it
The Moreau envelope \cite{Moreau,Yosida} of $f$ is the infimal convolution of $f$ with a quadratic penalty:
\begin{equation}\label{MELoss}
ME\left ( \textbf{x}_{1} \right ) =\min_{\textbf{x}_{2}}\left \{ f\left ( \textbf{x}_{2}\right)+\frac{\mu }{2}\left \| \textbf{x}_{2}-\textbf{x}_{1} \right \|^{2}_{2}\right\},
\end{equation}
where $\mu $ is a positive regularization parameter.
}

In a decentralized network, each node would communicate with neighbors and receive the transported parameter $\textbf{x}$ from neighbors. When node $i$ receives a transported parameter $\textbf{x}_{j}$ from its neighbor $j$, we formalize the Moreau envelope of node $i$'s expected local loss function $f_{i}$ following the transported parameter $\textbf{x}_{j}$ as follows:

\begin{equation}\label{DeMELoss}
ME_{i}\left ( \textbf{x}_{j} \right ) =\min_{\textbf{x}_{i}}\left \{ f_{i}\left ( \textbf{x}_{i}\right)+\frac{\mu }{2}\left \| \textbf{x}_{i}-\textbf{x}_{j} \right \|^{2}_{2}\right\}.
\end{equation}

\noindent
Summing up all the adjacency nodes of node $i$, we can get the Moreau envelope of node $i$'s local loss function in a decentralized network:
\begin{equation}\label{SUMDeMELoss}
\begin{aligned}
&&ME_{i}\left ( \textbf{X} \right )&=\frac{1}{K_{i}}  \sum_{j\in Adj_{i}} ME_{i}\left ( \textbf{x}_{j} \right )\\
&&\ &=\frac{1}{K_{i}} \sum_{j\in Adj_{i}}\min_{\textbf{x}_{i}}\left \{ f_{i}\left ( \textbf{x}_{i}\right)+\frac{\mu }{2}\left \| \textbf{x}_{i}-\textbf{x}_{j} \right \|^{2}_{2}\right\} ,
\end{aligned}
\end{equation}
where $Adj_{i}$ denotes the adjacency node set of node $i$, $K_{i}$ is the cardinality of $Adj_{i}$ and $\textbf{X}$ is the set of all parameters. The quadratic penalty $\frac{\mu }{2}\left \| \cdot  \right \| ^{2}_{2}$ of $ME_{i}$ can be generalized to an appropriate strongly convex function, while preserving properties such as continuously differentiable \cite{GeneralME,StronglyConvexME}. Therefore, we generalize Eq.(\ref{SUMDeMELoss}) as follows:
\begin{equation}\label{GenSUMDeMELoss}
DME_{i}\left ( \textbf{X} \right ) =\frac{1}{K_{i}} \sum_{j\in Adj_{i}}\min_{\textbf{x}_{i}}\left \{ f_{i}\left ( \textbf{x}_{i}\right)+\mathcal{C}\left ( \left \| \textbf{x}_{i}-\textbf{x}_{j} \right \|^{2}_{2} \right )  \right\},
\end{equation}
where $\mathcal{C}\left (\left \| \cdot \right \|_{2}^{2}    \right ) $ is a continuously differentiable and strongly convex function. We call $DME_{i}$ as the \textbf{Decentralized Moreau Envelope} of node $i$'s local loss function $f_{i}$. Replacing the expected local loss function $f_{i}$ in Problem 1 with $DME_{i}$, we can apply the decentralized Moreau envelope to decentralized optimization and have the following optimization problem.

\noindent
{\bf Remark 2} { \it
In a decentralized network with the decentralized Moreau envelope, the objective is to solve the following problem, with a two-stage of minimization involving both the individual nodes and the network.
\begin{equation*}\label{DMELoss}
\min_{\textbf{X}} \left \{ \frac{1}{N}\sum_{i=1}^{N}DME_{i}\left ( \textbf{X} \right ) \right \},\ \mathrm{s.t.}\ \textbf{x}_{i}=\textbf{x}_{j}\ \mathrm{for}\ \forall \textbf{x}_{i},\textbf{x}_{j}\in \textbf{X}.
\end{equation*}
}

However, the cost of locally optimizing $DME_{i}$ at node $i$ is unacceptable since each optimization has to perform the minimization process $K_{i}$ times. The Catalyst acceleration approach \cite{MoreauSmooth} proposed the fixed budget strategy, which fixes the number of local iterations to control the node-stage optimization price. Following the Catalyst acceleration approach, we fix the number of local iterations to one, adopt gradient descent with step size $\gamma$ at time $t$ to the node-stage optimization, and formalize it as follows:

\begin{equation}\label{DMEGD}
\begin{aligned}
&\textbf{x}^{(t)}_{i}-\textbf{x}^{(t-1)}_{i}\\
&=-\frac{\gamma}{K_{i}} \sum_{j\in Adj_{i}} \frac{\partial}{\partial \textbf{x}_{i}}  \left [ f_{i}\left ( \textbf{x}^{(t-1)}_{i}\right)+ \mathcal{C}\left ( \left \| \textbf{x}^{(t-1)}_{i}-\textbf{x}^{(t-1)}_{j} \right \|^{2}_{2} \right ) \right ] \\
&=-\gamma\frac{\partial}{\partial \textbf{x}_{i}} f_{i}\left ( \textbf{x}^{(t-1)}_{i}\right)\\
&\quad\ -\frac{\gamma}{K_{i}} \sum_{j\in Adj_{i}}\frac{\partial}{\partial \textbf{x}_{i}}\mathcal{C}\left ( \left \| \textbf{x}^{(t-1)}_{i}-\textbf{x}^{(t-1)}_{j} \right \|^{2}_{2} \right ).
\end{aligned}
\end{equation}

\noindent
We observe that the perturbation of node-stage optimization can be divided into a gradient of $f_{i}$ and a gradient of the sum of $\mathcal{C}\left (\cdot   \right ) $, where the former one is the local gradient, and the latter one is focusing on optimizing the distance between the local parameter and the adjacency parameters. Let $\mathcal{C}'\left ( \textbf{x} \right ) = \frac{\partial}{\partial \textbf{x}}\mathcal{C}\left ( \textbf{x} \right )$, we consider the gradient descent of the sum of $\mathcal{C}\left (\cdot   \right ) $ solely:

\begin{eqnarray}\label{ConvexCom}
&&\textbf{x}^{(t)}_{i}\nonumber\\
&&=\textbf{x}^{(t-1)}_{i}-\frac{\gamma}{K_{i}}\sum_{j\in Adj_{i}} \frac{\partial}{\partial \textbf{x}_{i}}\mathcal{C}\left ( \left \| \textbf{x}^{(t-1)}_{i}-\textbf{x}^{(t-1)}_{j} \right \|^{2}_{2} \right )\nonumber\\
&&=\textbf{x}^{(t-1)}_{i}\nonumber\\
&&\quad -\frac{\gamma}{K_{i}}\sum_{j\in Adj_{i}} \mathcal{C}'\left ( \left \| \textbf{x}^{(t-1)}_{i}-\textbf{x}^{(t-1)}_{j} \right \|^{2}_{2} \right )\frac{\partial \left \| \textbf{x}_{i}-\textbf{x}_{j} \right \|^{2}_{2}}{\partial\textbf{x}_{i} } \nonumber\\
&&=\left ( 1-\frac{2\gamma}{K_{i}} \sum_{j\in Adj_{i},j\ne i}\mathcal{C}'\left ( \left \| \textbf{x}^{(t-1)}_{i}-\textbf{x}^{(t-1)}_{j} \right \|^{2}_{2} \right ) \right )\textbf{x}^{(t-1)}_{i}\nonumber \\
&&\quad+\frac{2\gamma}{K_{i}} \sum_{j\in Adj_{i},j\ne i}\mathcal{C}'\left ( \left \| \textbf{x}^{(t-1)}_{i}-\textbf{x}^{(t-1)}_{j} \right \|^{2}_{2} \right ) \textbf{x}^{(t-1)}_{j} \nonumber\\
&&=\sum_{j\in Adj_{i}} \varsigma_{i,j}^{(t)} \textbf{x}^{(t-1)}_{j},
\end{eqnarray}

\noindent where

\begin{equation}\label{varsigmaPS}
\varsigma_{i,j}^{(t)}=\left\{
\begin{aligned}
&1-\frac{2\gamma}{K_{i}} \sum_{s\in Adj_{i},s\ne i}\mathcal{C}'\left ( \left \| \textbf{x}^{(t-1)}_{i}-\textbf{x}^{(t-1)}_{s} \right \|^{2}_{2} \right )\\
&\quad\quad\quad\quad\quad\quad\quad\quad\quad\quad\quad\quad\quad \mathrm{if} \ j=i,  \\
&\frac{2\gamma}{K_{i}} \mathcal{C}'\left ( \left \| \textbf{x}^{(t-1)}_{i}-\textbf{x}^{(t-1)}_{j} \right \|^{2}_{2} \right )\\
&\quad\quad\quad\quad\quad\quad\quad\  \mathrm{if}\ j\in Adj_{i}\ \mathrm{and}\ j\ne  i,\\
&0\quad\quad\quad\quad\quad\quad\quad\quad\quad\quad\quad\ \  othereise.
\end{aligned}
\right.
\end{equation}
Notice that with an appropriately chosen $\mathcal{C}'\left (\cdot   \right ) $, each coefficient $\varsigma_{i,j}^{(t)}$ could be non-negative, while the affine combination in Eq.(\ref{ConvexCom}) would become a convex combination and can be developed into a weighting method.

To develop the Moreau weighting method in Eq.(\ref{wPS}) from the convex combination Eq.(\ref{varsigmaPS}), we make three adaptations. First, we define the adjacency node set $Adj_i$ as the out-neighbor set $\left \{ s|(i,s)\in E^{(t)} \right \} $ to meet the protocol's property. Second, we use the buffer parameter $\textbf{x}_{i,j}^{(buff)}$ to approximate the parameter $\textbf{x}_{j}^{(t-1)}$ in Eq.~(\ref{varsigmaPS}). Since node $i$ may not have received $\textbf{x}_{j}^{(t-1)}$ in the last communication, we use the previously stored parameter $\textbf{x}_{i,j}^{(buff)}$ in node $i$'s buffer to approximate it. Third, $\mathcal{C}'\left( \left\| \cdot \right\|_{2}^{2} \right)$ is the derivative of the function $\mathcal{C}\left( \left\| \cdot \right\|_{2}^{2} \right)$ in Eq.~(\ref{GenSUMDeMELoss}), where $\mathcal{C}\left( \left\| \cdot \right\|_{2}^{2} \right)$ is continuously differentiable and strongly convex. Finding a perfect $\mathcal{C}'\left( \left\| \cdot \right\|_{2}^{2} \right)$ that meets all the requirements is quite challenging. Therefore, we simplify the requirements such that $\mathcal{C}'\left( \left\| \cdot \right\|_{2}^{2} \right)$ should be differentiable and strictly increasing, and we employ the negative exponential function in Eq.~(\ref{C}) for our Moreau weighting method.

In summary, Eqs.~(\ref{DMEGD}) and (\ref{ConvexCom}) show that with an appropriately chosen $\mathcal{C}'\left( \cdot \right)$, the optimization of $DME_{i}$ can be decomposed into a local gradient descent and a convex combination. Given that the main components of a decentralized protocol can be approximately viewed as a local gradient descent and an approximate convex combination, we incorporate the convex combination in the optimization of $DME_{i}$ into the weighted average of the Adaptive Weighting Push-SUM protocol, thereby proposing the Moreau weighting method. We verify the efficiency of the Moreau weighting method in handling statistical diversity by experiments.
\section{Experiments}
The experiments are implemented using Python 3.8 and CUDA 11.7. In the experiments, we adopt \textit{Resnet-18} \cite{Resnet} and \textit{Resnet-50} \cite{Resnet} as our models, where their parameters are the model parameter $\textbf{x}$. For the local data distribution $D_{i}$, we employ \textit{CIFAR-10} \cite{Cifar} and \textit{CIFAR-100} \cite{Cifar} as our datasets and control the network's statistical diversity by allocating a full or fixed split dataset to each node.

We compare five decentralized algorithms in the experiments, where three of them are based on the Push-SUM protocol and the other two of them are based on the Adaptive Weighting Push-SUM protocol. 

\noindent
{\bf Push-SUM Algorithms} {
\begin{itemize}
\item \texttt{SGP} \cite{SGP}: SGD with Push-SUM.
\item \texttt{MSGP}: Momentum SGD with Push-SUM.
\item \texttt{S-ADDOPT} \cite{SGTSGP}: Gradient Tracking  with Push-SUM.
\end{itemize}
}

\noindent
{\bf Adaptive Weighting Push-SUM Algorithms} {
\begin{itemize}
\item \texttt{SGAP (ours, Algorithm 2)}: SGD with Adaptive Weighting Push-SUM.
\item \texttt{MSGAP (ours, Algorithm 2)}: Momentum SGD with Adaptive Weighting Push-SUM.
\end{itemize}
}

\subsection{Wall-Clock Time}

\begin{table*}[htbp]
\caption{The final test accuracy and loss of experiments.}
\centering
\begin{tabular}{|cc|cccc|cccc|cc|}
\hline
\multicolumn{2}{|c|}{Model + Dataset}                       & \multicolumn{4}{c|}{Resnet18 + Cifar10}                                                                                          & \multicolumn{4}{c|}{Resnet50 + Cifar100}                                                                                         & \multicolumn{2}{c|}{Average}                         \\ \hline
\multicolumn{2}{|c|}{Topology}                              & \multicolumn{1}{c|}{Full}           & \multicolumn{1}{c|}{Divide}         & \multicolumn{1}{c|}{Exp}            & Random         & \multicolumn{1}{c|}{Full}           & \multicolumn{1}{c|}{Divide}         & \multicolumn{1}{c|}{Exp}            & Random         & \multicolumn{1}{c|}{R18 + C10}      & R50 + C100     \\ \hline
\multicolumn{1}{|c|}{\multirow{2}{*}{SGP}}      & Acc(\%)   & \multicolumn{1}{c|}{59.31}          & \multicolumn{1}{c|}{62.19}          & \multicolumn{1}{c|}{61.33}          & 53.48          & \multicolumn{1}{c|}{51.28}          & \multicolumn{1}{c|}{51.97}          & \multicolumn{1}{c|}{52.96}          & 47.33          & \multicolumn{1}{c|}{59.08}          & 50.89          \\ \cline{2-12} 
\multicolumn{1}{|c|}{}                          & Loss(E-3) & \multicolumn{1}{c|}{14.29}          & \multicolumn{1}{c|}{13.68}          & \multicolumn{1}{c|}{13.97}          & 16.24          & \multicolumn{1}{c|}{24.29}          & \multicolumn{1}{c|}{23.21}          & \multicolumn{1}{c|}{23.43}          & 25.24          & \multicolumn{1}{c|}{14.55}          & 24.04          \\ \hline
\multicolumn{1}{|c|}{\multirow{2}{*}{MSGP}}     & Acc(\%)   & \multicolumn{1}{c|}{63.39}          & \multicolumn{1}{c|}{62.72}          & \multicolumn{1}{c|}{63.26}          & 55.38          & \multicolumn{1}{c|}{36.03}          & \multicolumn{1}{c|}{44.03}          & \multicolumn{1}{c|}{47.69}          & 38.69          & \multicolumn{1}{c|}{61.19}          & 41.61          \\ \cline{2-12} 
\multicolumn{1}{|c|}{}                          & Loss(E-3) & \multicolumn{1}{c|}{14.75}          & \multicolumn{1}{c|}{15.52}          & \multicolumn{1}{c|}{15.56}          & 17.48          & \multicolumn{1}{c|}{45.73}          & \multicolumn{1}{c|}{36.49}          & \multicolumn{1}{c|}{28.09}          & 42.34          & \multicolumn{1}{c|}{15.83}          & 38.16          \\ \hline
\multicolumn{1}{|c|}{\multirow{2}{*}{SGAP}}     & Acc(\%)   & \multicolumn{1}{c|}{62.97}          & \multicolumn{1}{c|}{63.25}          & \multicolumn{1}{c|}{62.36}          & \textbf{61.09} & \multicolumn{1}{c|}{\textbf{56.10}} & \multicolumn{1}{c|}{\textbf{57.06}} & \multicolumn{1}{c|}{\textbf{56.93}} & \textbf{49.93} & \multicolumn{1}{c|}{62.42}          & \textbf{55.00} \\ \cline{2-12} 
\multicolumn{1}{|c|}{}                          & Loss(E-3) & \multicolumn{1}{c|}{13.27}          & \multicolumn{1}{c|}{12.51}          & \multicolumn{1}{c|}{\textbf{13.20}} & \textbf{13.44} & \multicolumn{1}{c|}{\textbf{22.18}} & \multicolumn{1}{c|}{\textbf{20.74}} & \multicolumn{1}{c|}{\textbf{21.68}} & \textbf{24.24} & \multicolumn{1}{c|}{\textbf{13.11}} & \textbf{22.21} \\ \hline
\multicolumn{1}{|c|}{\multirow{2}{*}{MSGAP}}    & Acc(\%)   & \multicolumn{1}{c|}{\textbf{66.58}} & \multicolumn{1}{c|}{\textbf{68.62}} & \multicolumn{1}{c|}{\textbf{64.52}} & 58.83          & \multicolumn{1}{c|}{50.11}          & \multicolumn{1}{c|}{50.77}          & \multicolumn{1}{c|}{51.23}          & 46.79          & \multicolumn{1}{c|}{\textbf{64.64}} & 49.73          \\ \cline{2-12} 
\multicolumn{1}{|c|}{}                          & Loss(E-3) & \multicolumn{1}{c|}{\textbf{12.85}} & \multicolumn{1}{c|}{\textbf{11.33}} & \multicolumn{1}{c|}{15.29}          & 16.89          & \multicolumn{1}{c|}{28.08}          & \multicolumn{1}{c|}{33.87}          & \multicolumn{1}{c|}{26.90}          & 32.11          & \multicolumn{1}{c|}{14.09}          & 30.24          \\ \hline
\multicolumn{1}{|c|}{\multirow{2}{*}{S-ADDOPT}} & Acc(\%)   & \multicolumn{1}{c|}{60.20}          & \multicolumn{1}{c|}{59.47}          & \multicolumn{1}{c|}{59.76}          & 54.36          & \multicolumn{1}{c|}{54.24}          & \multicolumn{1}{c|}{53.97}          & \multicolumn{1}{c|}{55.62}          & 31.36          & \multicolumn{1}{c|}{58.45}          & 48.80          \\ \cline{2-12} 
\multicolumn{1}{|c|}{}                          & Loss(E-3) & \multicolumn{1}{c|}{13.76}          & \multicolumn{1}{c|}{14.35}          & \multicolumn{1}{c|}{13.91}          & 14.96          & \multicolumn{1}{c|}{22.58}          & \multicolumn{1}{c|}{23.18}          & \multicolumn{1}{c|}{22.78}          & 27.36          & \multicolumn{1}{c|}{14.25}          & 23.98          \\ \hline
\end{tabular}
\label{ACCLOSS}
\end{table*}

\begin{figure}[htbp]
    \centering
    \includegraphics[width=\linewidth]{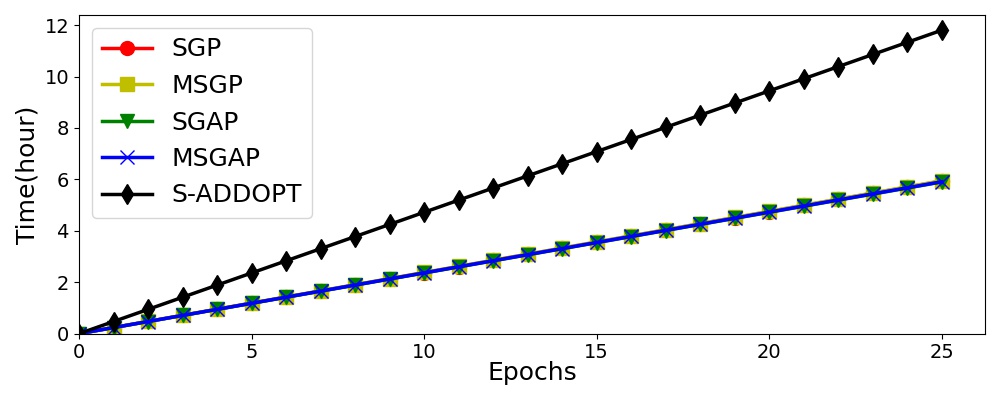}
    \caption{Cumulative Time Costs for Four Algorithms.}
    \label{process_time}
\end{figure}

The following experiments are implemented on a NVIDIA RTX 4090. The network is a fully connected network with scale $N = 6$. The bandwidth of each communication channel is $1Gbps$. The strongly connected period $B = 1$. We train the ResNet-18 on the CIFAR-10 dataset, while each node has the whole dataset with batch size 100. The wall clock time cost by the five algorithms in the training is shown in Fig. \ref{process_time}.

The results shown in Fig. \ref{process_time} indicate that the parallel strategy we presented in Fig. \ref{process_graph} is effective, as it fully eliminates the computational delay of the Adaptive Weighting Push-SUM protocol compared to the Push-SUM protocol when executing the Moreau weighting method under a bandwidth of $1Gbps$. 


In Fig. \ref{process_time}, the total time cost for SGP, MSGP, SGAP, and MSGAP is approximately $5.9$ hours, whereas for SADDOPT it is approximately $11.8$ hours. The results further indicate the significance of communication overhead in decentralized optimization. Although Momentum SGD computes an additional momentum term $\textbf{m}$ compared to SGD, the time cost of both \texttt{SGP} versus \texttt{MSGP} and \texttt{SGAP} versus \texttt{MSGAP} does not exhibit a notable difference. By contrast, the \texttt{S-ADDOPT} algorithm, which employs gradient tracking, has doubled the time cost. This disparity is because \texttt{S-ADDOPT} needs to communicate two $d$-dimensional vectors, which are the model parameter and the gradient tracking term, whereas the other four algorithms only need to communicate one $d$-dimensional vector, which is the model parameter.

\subsection{Statistical Diversity}

The following experiments are implemented on a server with 6 NVIDIA A100. The network scale $N=6$. We train the ResNet-18 model on the CIFAR-10 dataset with $\gamma = 0.01$ and the ResNet-50 model on the CIFAR-100 dataset with $\gamma = 0.1$. Both models were trained with $\beta = 0.8$, batch size of each node is $100$.

For statistical diversity, we first divide all nodes in the network into two clusters where nodes $0\sim 2$ are cluster 0, and nodes $3\sim 5$ are cluster 1. Then, for the CIFAR-10/CIFAR-100 dataset, each node in cluster 0 only has train data with label $0\sim 4$/$0\sim 49$, while each node in cluster 1 only has train data with label $5\sim 9$/$50\sim 99$. To get test accuracy, we all-reduce every model in the network to get the average parameter $\overline{\textbf{x}}$ and then test it on the complete test dataset of CIFAR-10/CIFAR-100. Our dataset splits are fixed and clear since we hope our experimental results can be easily reproduced.

\begin{figure}[htbp]
    \centering
    \includegraphics[width=\linewidth]{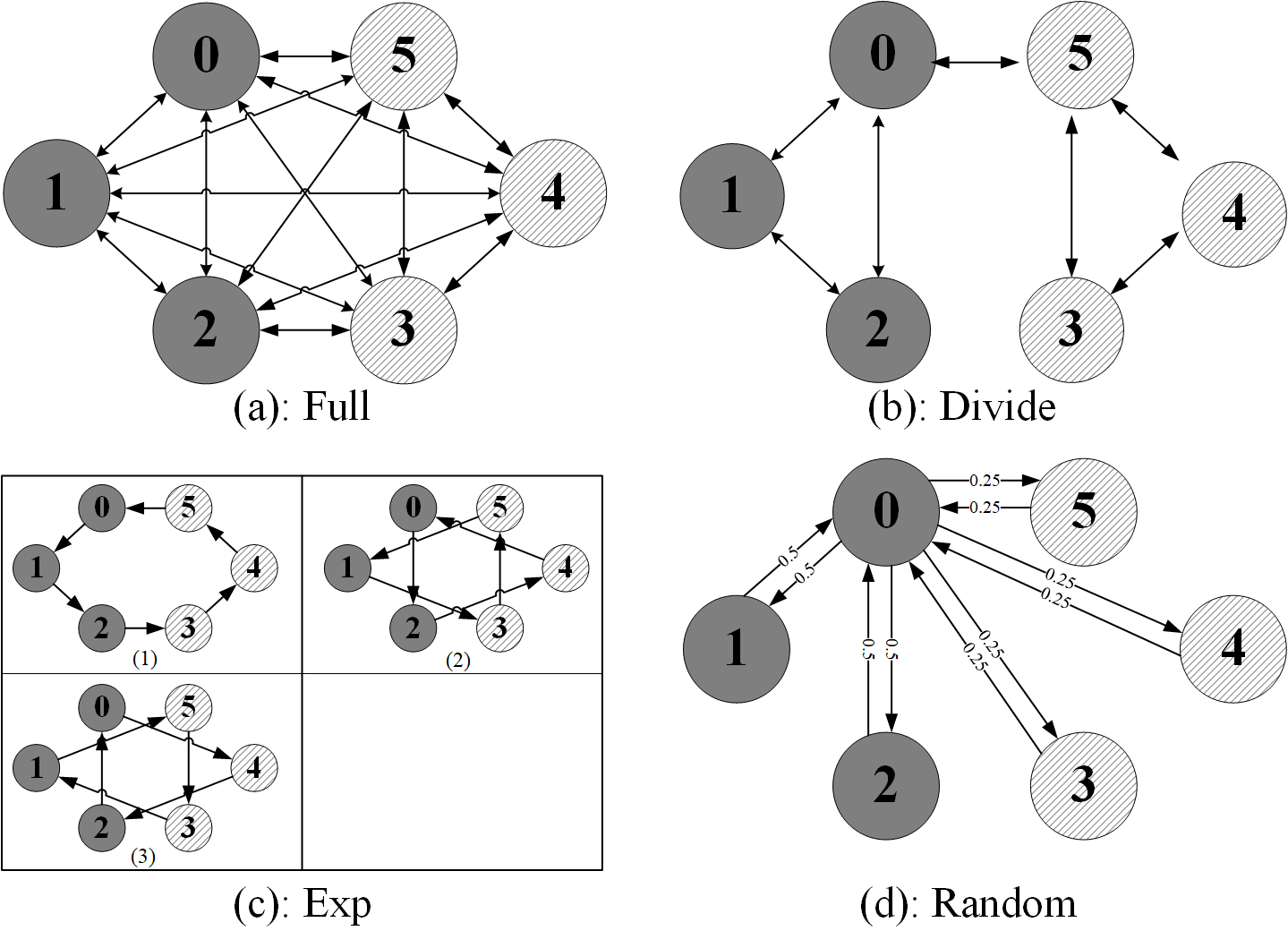}
    \caption{The Four Network Topologies.}
    \label{topo}
\end{figure}

Moreover, to investigate the performance of the protocols under different $\Delta$ and $B$ in Assumption 1, we conducted experiments under four different network topologies. These topologies are illustrated in Fig. \ref{topo} and described as follows.

\begin{itemize}
    \item {\bf Full:} A fully connected graph.
    \item {\bf Divide:} Each cluster are fully connected inside, but only has one two-way channel between the two clusters.
    \item {\bf Exp:} This topology is proposed by \cite{SGP,AsynSGP}, each node cyclically sends its parameters to neighbors. Fig. \ref{topo}(c) shows the three subgraphs used cyclically in Exp.
    \item {\bf Random:} This topology is based on a fully connected graph, but each one-way channel is opened randomly. At the last epoch of each strongly connected period, all channels are forced open. The opening probability is set to 0.5 for inner-cluster channels, 0.25 for inter-cluster channels, and the strongly connected period $B=8$. Fig. \ref{topo}(d) only shows node $0$'s connection.
    
    
\end{itemize}

Table \ref{ACCLOSS} shows the final test accuracy and loss for the five algorithms after 25 epochs of training.

For the ResNet-18 model on the CIFAR-10 dataset with statistical diversity. Our \texttt{SGAP}, compared to \texttt{SGP}, has improved the average final accuracy absolutely by $3.34$ and relatively by $5.65\%$, while reduced the average final loss absolutely by $1.44$ and relatively by $9.90\%$. Our \texttt{MSGAP}, compared to \texttt{MSGP}, has improved the average final accuracy absolutely by $3.45$ and relatively by $5.64\%$, while reduced the average final loss absolutely by $1.74$ and relatively by $10.99\%$.

For the ResNet-50 model on the CIFAR-100 dataset with statistical diversity. Our \texttt{SGAP}, compared to \texttt{SGP}, has improved the average final accuracy absolutely by $4.11$ and relatively by $8.08\%$, while reduced the average final loss absolutely by $1.83$ and relatively by $7.61\%$. Our \texttt{MSGAP}, compared to \texttt{MSGP}, has improved the average final accuracy absolutely by $8.12$ and relatively by $19.51\%$, while reduced the average final loss absolutely by $7.92$ and relatively by $20.75\%$.


\texttt{S-ADDOPT} is an effective solution for statistical diversity \cite{NonIIDGT}. Our \texttt{SGAP} outperforms \texttt{S-ADDOPT} in all experiments, while \texttt{MSGAP} is competitive with it.


Based on the above experimental results, we empirically verify that the Adaptive Weighting Push-SUM protocol with Moreau weighting is more efficient than the original Push-SUM protocol for statistical diversity.
\section{Conclusion}
In this paper, we proposed the Adaptive Weighting Push-SUM protocol, adopted SGD and Momentum SGD to develop \texttt{SGAP} and \texttt{MSGAP}, and provided the Moreau weighting method to meet our protocol's weighting requirements. We provided theoretical analysis and comparisons for our protocol and algorithms. For the protocol, we reduced the distance upper bound from $O(1)$ to $O(1/N)$. For the algorithms, we reduced the upper bound in handling statistical diversity from $O(Nd/T)$ to $O(N/T)$, where $d$ represents the parameter size of the training model and $d$ is far beyond millions for popular DNN models. We proposed the Moreau weighting method for the flexible weight matrix definitions in our protocol and empirically verified that our protocol with Moreau weighting is more efficient than the Push-SUM protocol in handling statistical diversity in decentralized optimization.
\appendices
\section*{Notation}
\noindent
{\bf Fact 1} {\it 
For arbitrary matrix $A$, the following formula holds
\begin{equation}
\left \| \textbf{A} \right \|_{F}^{2}=\sum_{i=1}^{N}\left \| \textbf{a}_{i} \right \|^{2}_{2}
\end{equation}
where $\textbf{a}_{i}$ is $\textbf{A}'s$ i'th row(column) vector, $N$ is $\textbf{A}'s$ row(column) dimension.
}

\noindent
{\bf Fact 2} {\it 
Let $\textbf{A}$ be $M\times N$ arbitrary matrix, $\textbf{a}$ be N-dimensional arbitrary vector,then we have
\begin{equation}
\left \| \textbf{A}\textbf{a}  \right \|_{2}\le \left \| \textbf{A} \right \|_{F}\left \| \textbf{a} \right \|_{2} 
\end{equation}
}

\section{Proof of Theorem 2}

From Theorem 1 of the Adaptive Weighting Push-SUM protocol, we can get the following lemma for \texttt{MSGAP}.

\noindent
{\bf Lemma 3} {\it 
Suppose that Assumption 1, 2, 3 hold, under Definition 1, consider sequence $\{\textbf{y}_{i}^{(t)}\}_{t\geq0}$ and $\{\overline{\textbf{x}}^{(t)}\}_{t\geq0}$, under $\gamma<\frac{(1-\alpha)^{2}}{12CNL}$ following established for all $T>\Delta B$.
\begin{eqnarray*}\label{big_lem1_equ}
&&\sum_{t=0}^{T-1}\sum_{i=1}^{N}\mathbb{E}\left [ \left \| \textbf{y}_{i}^{(t)}-\overline{\textbf{x}}^{(t)}\right \|^{2}_{2}\right ]\nonumber\\
&&\le\frac{1}{Q}\left(2N+\frac{6C^{2}N^{2}}{1-\lambda ^{2}}\right) \left \| \textbf{X}^{(0)} \right \|^{2}_{F}\nonumber\\
&&\quad +72\frac{\gamma^{2} C^{2}N^{3}}{Q(1-\alpha)^{4}}\sum_{t=0}^{T-1}\mathbb{E}\left [ \left \| \frac{1}{N}\sum_{i=1}^{N}\triangledown f_{i}(\textbf{y}_{i}^{(t)})\right \|^{2}_{2}\right ]\nonumber\\
&&\quad +9\frac{\gamma^{2} C^{2}N}{Q(1-\alpha)^{4}}\left ( T+N^{2}\Delta B \right ) \sigma^{2}\nonumber\\
&&\quad +36\frac{\gamma^{2} C^{2}N}{Q(1-\alpha)^{4}}\left ( T+N^{2}\Delta B \right )\kappa ^{2},
\end{eqnarray*}

\noindent
where $Q:=1-144\frac{\gamma^{2} C^{2}N^{2}L^{2}}{(1-\alpha)^{4}}>0$
}

For the average gradient of \texttt{MSGAP}, by Assumption 3, we can get the following lemma.

\noindent
{\bf Lemma 4} {\it 
Suppose that Assumption 1 holds, for $\textbf{g}_{i}^{(t)}$ in Algorithm 2, we have the
following inequality.
\begin{equation*}\label{lemma_grad_upper}
\mathbb{E}\left [  \left \| \frac{1}{N}\sum_{i=1}^{N}\textbf{g}_{i}^{(t)} \right \|^{2}_{2}\right ] \le \frac{1}{N}\sigma^{2}+\mathbb{E}\left [  \left \| \frac{1}{N}\sum_{i=1}^{N}\triangledown f_{i}(\textbf{y}_{i}^{(t)}) \right \|^{2}_{2}\right ].
\end{equation*}
}

From the theoretical analysis of Yu \textit{et al.} \cite{Momentum1}, we define the following auxiliary sequence $\{\overline{\textbf{z}}^{(t)}\}_{t \geq 0}$ for our \texttt{MSGAP}.

\noindent
{\bf Definition 4} {\it
The sequence $\{\overline{\textbf{z}}^{(t)}\}_{t \geq 0}$ is defined as follows:
\begin{eqnarray*}\label{aux_def_equ}
\overline{\textbf{z}}^{(t)} :=\left\{\begin{matrix}
\overline{\textbf{x}}^{(t)} &&t=0,\\
\frac{1}{1-\beta}\overline{\textbf{x}}^{(t)}-\frac{\beta}{1-\beta}\overline{\textbf{x}}^{(t-1)}&&t\geq 1.
\end{matrix}\right.
\end{eqnarray*}
}

\noindent
{\bf Lemma 5} {\it 
Consider the sequence $\{\overline{\textbf{z}}^{(t)}\}_{t\geq 0}$, \texttt{MSGAP} ensures the following equality holds for all $t\geq0$
\begin{equation*}
\overline{\textbf{z}}^{(t+1)}-\overline{\textbf{z}}^{(t)}=-\frac{\gamma}{1-\beta}\frac{1}{N}\sum_{i=1}^{N}\textbf{g}_{i}^{(t)}.
\end{equation*}
}

\noindent
{\bf Lemma 6} {\it 
Consider sequences $\{\overline{\textbf{x}}^{(t)}\}_{t\geq 0}$ and $\{\overline{\textbf{z}}^{(t)}\}_{t\geq 0}$, \texttt{MSGAP} ensures the following inequality holds for all $T\geq1$
\begin{equation*}
\sum_{t=0}^{T-1}\left \|\overline{\textbf{z}}^{(t)}-\overline{\textbf{x}}^{(t)}\right \|^{2}_{2}\leq \frac{\gamma^{2}\beta^{2}}{(1-\beta)^{4}}\sum_{t=0}^{T-1}\left \|\frac{1}{N}\sum_{i=1}^{N}\textbf{g}_{i}^{(t)}\right \|^{2}_{2}.
\end{equation*}
}

All proof of lemmas can be found in the appendix later.\bigskip

\noindent
\textbf{Main Proof of Theorem 2}

\begin{proof}
Fix $t\geq 0$. By Assumption 2, we have
\begin{eqnarray}\label{op1_the_pro_1}
\mathbb{E}\left [ f(\overline{\textbf{z}}^{(t+1)} ) \right ]&\leq& \mathbb{E}\left [ f(\overline{\textbf{z}}^{(t)} ) \right ]\nonumber  \\
&&+\mathbb{E}\left [ \left \langle \triangledown f(\overline{\textbf{z}}^{(t)}),\overline{\textbf{z}}^{(t+1)}-\overline{\textbf{z}}^{(t)} \right \rangle  \right ]\nonumber \\
&&+\frac{L}{2}\mathbb{E}\left [ \left \| \overline{\textbf{z}}^{(t+1)}-\overline{\textbf{z}}^{(t)} \right \|^{2}_{2}  \right ]   .
\end{eqnarray}

\noindent
For the second term on the right side of \eqref{op1_the_pro_1}
\begin{equation}\label{op1_the_pro_2}
\begin{aligned}
&\mathbb{E}\left [ \left \langle \triangledown f(\overline{\textbf{z}}^{(t)}),\overline{\textbf{z}}^{(t+1)}-\overline{\textbf{z}}^{(t)} \right \rangle  \right ]\\
&\overset{(a)}{=}-\frac{\gamma}{1-\beta}\mathbb{E}\left [ \left \langle \triangledown f(\overline{\textbf{z}}^{(t)}),\frac{1}{N}\sum_{i=1}^{N}\textbf{g}_{i}^{(t)} \right \rangle  \right ] \\
&\overset{(b)}{=}-\frac{\gamma}{1-\beta}\mathbb{E}\left [ \left \langle \triangledown f(\overline{\textbf{z}}^{(t)}),\frac{1}{N}\sum_{i=1}^{N}\triangledown f_{i}(\textbf{y}^{(t)}_{i}) \right \rangle  \right ]\\
&=\underbrace{-\frac{\gamma}{1-\beta}\mathbb{E}\left [ \left \langle \triangledown f(\overline{\textbf{z}}^{(t)})-\triangledown f(\overline{\textbf{x}}^{(t)}),\frac{1}{N}\sum_{i=1}^{N}\triangledown f_{i}(\textbf{y}^{(t)}_{i}) \right \rangle  \right ]}_{\textbf{a}} \\
&\quad -\frac{\gamma}{1-\beta}\mathbb{E}\left [\underbrace{ \left \langle \triangledown f(\overline{\textbf{x}}^{(t)}),\frac{1}{N}\sum_{i=1}^{N}\triangledown f_{i}(\textbf{y}^{(t)}_{i}) \right \rangle}_{\textbf{b}}  \right ].
\end{aligned}
\end{equation}

\noindent
(a): By Lemma 5.\\
\noindent
(b): $\overline{\textbf{z}}^{(t)}$ is determined by $\xi^{[t-1]}=\left [ \xi^{(0)},\xi^{(1)},...,\xi^{(t-1)} \right ] $, while $\textbf{g}_{i}^{(t)}$ is determined by $\xi^{(t)}$, so $\overline{\textbf{z}}^{(t)}$ and $\textbf{g}_{i}^{(t)}$ are independent.

\noindent
For $\textbf{a}$ in Eq.\eqref{op1_the_pro_2},
\begin{equation}\label{op1_the_pro_3}
\begin{aligned}
&-\frac{\gamma}{1-\beta}\mathbb{E}\left [ \left \langle \triangledown f(\overline{\textbf{z}}^{(t)})-\triangledown f(\overline{\textbf{x}}^{(t)}),\frac{1}{N}\sum_{i=1}^{N}\triangledown f_{i}(\textbf{y}^{(t)}_{i}) \right \rangle  \right ]\\
&\overset{(a)}{\leq}\frac{1-\beta}{2\beta L}\mathbb{E}\left [ \left \| \triangledown f(\overline{\textbf{z}}^{(t)})-\triangledown f(\overline{\textbf{x}}^{(t)}) \right \|^{2}_{2}\right ]\\
&\quad +\frac{\beta L\gamma^{2}}{2(1-\beta)^{3}}\mathbb{E}\left [ \left \| \frac{1}{N}\sum_{i=1}^{N}\triangledown f_{i}(\textbf{y}^{(t)}_{i}) \right \|^{2}_{2}\right ]\\
&\overset{(b)}{\leq}\frac{(1-\beta)L}{2\beta }\mathbb{E}\left [ \left \|\overline{\textbf{z}}^{(t)}-\overline{\textbf{x}}^{(t)}  \right \|^{2}_{2}\right ]\\
&\quad +\frac{\beta L\gamma^{2}}{2(1-\beta)^{3}}\mathbb{E}\left [ \left \| \frac{1}{N}\sum_{i=1}^{N}\triangledown f_{i}(\textbf{y}^{(t)}_{i}) \right \|^{2}_{2}\right ].
\end{aligned}
\end{equation}

\noindent
(a): By $\left \langle \textbf{a} ,\textbf{b} \right \rangle \leq \frac{1}{2}\left \| \textbf{a} \right \| ^{2}_2 +\frac{1}{2}\left \| \textbf{b} \right \| ^{2}_2$.\\
\noindent
(b) By the smoothness of $f(\cdot)$.

\noindent
For $\textbf{b}$ in Eq.\eqref{op1_the_pro_2},
\begin{equation}\label{op1_the_pro_4}
\begin{aligned}
& \left \langle \triangledown f(\overline{\textbf{x}}^{(t)}),\frac{1}{N}\sum_{i=1}^{N}\triangledown f_{i}(\textbf{y}^{(t)}_{i}) \right \rangle\\
&\overset{(a)}{\geq} \frac{1}{2} \left \| \triangledown f(\overline{\textbf{x}}^{(t)}) \right \|^{2}_{2} +\frac{1}{2}\left \| \frac{1}{N}\sum_{i=1}^{N}\triangledown f_{i}(\textbf{x}^{(t)}_{i}) \right \|^{2}_{2} \\
&\quad -\frac{L^{2}}{2}\frac{1}{N}\sum_{i=1}^{N} \left \|\overline{\textbf{x}}^{(t)}-\textbf{y}_{i}^{(t)}\right \|^{2}_{2}.
\end{aligned}
\end{equation}

\noindent
(a): By $\left \langle \textbf{a} ,\textbf{b} \right \rangle = \frac{1}{2}\left \| \textbf{a} \right \| ^{2}_2 +\frac{1}{2}\left \| \textbf{b} \right \| ^{2}_2-\frac{1}{2}\left \| \textbf{a}-\textbf{b} \right \| ^{2}_2$ and Assumption 2.

\noindent
Substitute Eq.\eqref{op1_the_pro_3} and Eq.\eqref{op1_the_pro_4} into Eq.\eqref{op1_the_pro_2}. Then, based on Lemma 5, applying it to Eq.\eqref{op1_the_pro_1}, we obtain
\begin{equation}\label{op1_the_pro_5}
\begin{aligned}
&\mathbb{E}\left [ f(\overline{\textbf{z}}^{(t+1)} ) \right ]\\
&\overset{(a)}{\leq} \mathbb{E}\left [ f(\overline{\textbf{z}}^{(t)} ) \right ]+\frac{(1-\beta)L}{2\beta }\mathbb{E}\left [ \left \|\overline{\textbf{z}}^{(t)}-\overline{\textbf{x}}^{(t)}  \right \|^{2}_{2}\right ]\\
&\quad+\left (\frac{\beta L\gamma^{2}}{2(1-\beta)^{3}}-\frac{\gamma}{2(1-\beta)}\right ) \mathbb{E}\left [ \left \| \frac{1}{N}\sum_{i=1}^{N}\triangledown f_{i}(\textbf{y}^{(t)}_{i}) \right \|^{2}_{2}\right ]\\
&\quad -\frac{\gamma}{2(1-\beta)}\mathbb{E}\left [  \left \| \triangledown f(\overline{\textbf{x}}^{(t)}) \right \|^{2}_{2}\right ] \\
&\quad +\frac{\gamma L^{2}}{2(1-\beta)}\frac{1}{N} \sum_{i=1}^{N}\mathbb{E}\left [ \left \|\overline{\textbf{x}}^{(t)}-\textbf{y}_{i}^{(t)}  \right \|^{2}_{2}  \right ]\\
&\quad +\frac{\gamma^{2}L}{2(1-\beta)^{2}}\mathbb{E}\left [  \left \| \frac{1}{N}\sum_{i=1}^{N}\textbf{g}_{i}^{(t)} \right \|^{2}_{2}\right ].
\end{aligned}
\end{equation}

\noindent
Rearrange \eqref{op1_the_pro_5}, sum over $t\in \{0,1,...,T-1\}$ and divide both sides by $\frac{\gamma}{2(1-\beta)}$, we have
\begin{equation}\label{op1_the_pro_7}
\begin{aligned}
&\sum_{t=0}^{T-1}\mathbb{E}\left [  \left \| \triangledown f(\overline{\textbf{x}}^{(t)}) \right \|^{2}_{2}\right ]\\
&\overset{(a)}{\leq} \frac{2(1-\beta)}{\gamma}\left ( \mathbb{E}\left [ f(\overline{\textbf{z}}^{(0)} ) \right ]-\mathbb{E}\left [ f(\overline{\textbf{z}}^{(T)} ) \right ]\right ) \\
&\quad-\left (Q-\frac{(1+\beta)L\gamma}{(1-\beta)^{2}}\right ) \sum_{t=0}^{T-1}\mathbb{E}\left [ \left \| \frac{1}{N}\sum_{i=1}^{N}\triangledown f_{i}(\textbf{y}^{(t)}_{i}) \right \|^{2}_{2}\right ]\\
&\quad+\frac{\gamma L}{N(1-\beta)^{2}}T\sigma^{2}+\left(4L^{2}+\frac{12C^{2}NL^{2}}{1-\lambda^{2}}\right)\left \| \textbf{X}^{(0)}\right \|^{2}_{F} \\
&\quad+\frac{\gamma^{2}C^{2}L^{2}}{(1-\alpha)^{4}}\left ( T+N^{2}\Delta B \right )\left ( 18\sigma^{2}+72\kappa ^{2} \right ) \\
&\overset{(b)}{\leq} \frac{2(1-\beta)}{\gamma}\left ( f(\overline{\textbf{x}}^{(0)} )- f^{*}\right )\\
&\quad+\left(4L^{2}+\frac{12C^{2}NL^{2}}{1-\lambda^{2}}\right)\left \| \textbf{X}^{(0)}\right \|^{2}_{F}+\frac{\gamma L}{N(1-\beta)^{2}}T\sigma^{2}\\
&\quad+\frac{\gamma^{2}C^{2}L^{2}}{(1-\alpha)^{4}}\left ( T+N^{2}\Delta B \right )\left ( 18\sigma^{2}+72\kappa ^{2} \right ).
\end{aligned}
\end{equation}

\noindent
(a): By Lemma 3, 4, and 6.\\
\noindent
(b): By $\gamma\leq min\left \{ \frac{(1-\alpha)^{2}}{12\sqrt{2}CNL},\frac{(1-\beta)^{2}}{2L(1+\beta)} \right \}$, $\overline{\textbf{z}}^{(0)}=\overline{\textbf{x}}^{(0)}$ in Definition 3, and $f^{*}$ is the minimum value in  Problem 1.

\noindent
Divide both sides by $T$, we can get Theorem 2.
\begin{equation}\label{op1_the_pro_8}
\begin{aligned}
&\frac{1}{T}\sum_{t=0}^{T-1}\mathbb{E}\left [ \left \| \triangledown f(\overline{\textbf{x}}^{(t)}) \right \|^{2}_{2} \right ]\nonumber \\
&\leq \frac{2-2\beta}{\gamma T}\left ( f(\overline{\textbf{x}}^{(0)} )- f^{*}\right )+\left(\frac{4L^{2}}{T}+\frac{12C^{2}NL^{2}}{T(1-\lambda^{2})}\right)\left \| \textbf{X}^{(0)}\right \|^{2}_{F}\\
&\quad+\frac{\gamma L}{N(1-\beta)^{2}}\sigma^{2}+\frac{18\gamma^{2}C^{2}L^{2}}{(1-\alpha)^{4}}\left ( 1+\frac{N^{2}\Delta B}{T} \right ) \sigma^{2}\\
&\quad +\frac{72\gamma^{2}C^{2}L^{2}}{(1-\alpha)^{4}}\left ( 1+\frac{N^{2}\Delta B}{T} \right ) \kappa ^{2}.
\end{aligned}
\end{equation}
\end{proof}
\clearpage
\onecolumn

\section{Proof of Lemma 1}
\noindent
{\bf Lemma 7} {\it
Suppose that the graph sequence $\{\mathcal{G}^{(t)}\}$ satisfies Assumption 1. Then for the weighting matrix sequence $\{\textbf{W}^{(t)}\}$ generated by $\{\mathcal{G}^{(t)}\}$ and each integer $s\geq 0$, there exists a stochastic vector $\phi(s)$ such that for all $i,j$ and $t\geq s$,
\begin{equation}\label{lem_w_exi_equ}
\left | \left [ (\textbf{W}^{(t)})^{T}...(\textbf{W}^{(s+1)})^{T}(\textbf{W}^{(s)}) ^{T}\right ]_{i,j}-\phi_{j}(s) \right |\leq k\lambda ^{t-s}
\end{equation}
for some $k$ and $\lambda\in(0,1)$, where $\left [ \textbf{A}\right ]_{i,j}$ denotes the entry of matrix $\textbf{A}$ at the $i$-th row and $j$-th column. 
}

\noindent
Lemma 1 is widely used in the theoretical analyses of the Push-SUM protocol works \cite{PerturbedPS}. See (\cite{lemma3/1,lemma3/2,lemma3/3,lemma3/4,lemma3/5,PerturbedPS}) for proofs and similar statements of Lemma 7.

\noindent
{\bf Lemma 8} {\it
Suppose that the graph sequence $\{\mathcal{G}^{(t)}\}$ satisfies Assumption 1. Then for the weighting matrix sequence $\{\textbf{W}^{(t)}\}$ generated by $\{\mathcal{G}^{(t)}\}$, Lemma 7 is satisfied with
\begin{equation}\label{lem_w_rate_equ}
k=2,\quad \lambda=(1-\delta^{\Delta B})^{\frac{1}{\Delta B}}
\end{equation}
}

\begin{proof}
For an arbitrary vector $\textbf{x}(s)$, let $\textbf{x}(t):=(\textbf{W}^{(t)})^{T} \cdots (\textbf{W}^{(s)})^{T}\textbf{x}(s)$ for all $t > s \geq 0$. When $t-s$ is divided by $\Delta B$, any arbitrary sub-sequence of $(\textbf{W}^{(t)})^{T} \cdots (\textbf{W}^{(s)})^{T}$ with length $B$ can generate a strongly connected graph with a diameter of at most $\Delta$, according to Assumption 1.

Therefore, any arbitrary sub-sequence of $(\textbf{W}^{(t)})^{T} \cdots (\textbf{W}^{(s)})^{T}$ with length $\Delta B$ can generate a complete graph, and the product of this sub-sequence has no zero elements:
\begin{equation}\label{lemma4/1}
\textbf{W}[i+\Delta B-1:i]=(\textbf{W}^{(i+\Delta B-1)})^{T} \cdots (\textbf{W}^{(i)})^{T}, \quad \forall i \geq 0
\end{equation}

Meanwhile, the maximum element of $\textbf{W}[i+\Delta B-1:i]$ is at most 1, and the minimum is at least $\delta^{\Delta B}$. Since each $(\textbf{W}^{(i)})^{T}$ is row stochastic, we have
\begin{equation}\label{lemma4/2}
\max_{1 \leq i \leq N} \textbf{x}_{i}(t) - \min_{1 \leq i \leq N} \textbf{x}_{i}(t) \leq \left(1 - \delta^{\Delta B}\right)^{\left\lfloor (t-s)/\Delta B \right\rfloor} \left(\max_{1 \leq i \leq N} \textbf{x}_{i}(s) - \min_{1 \leq i \leq N} \textbf{x}_{i}(s)\right)
\end{equation}

It should be noted that (\ref{lemma4/2}) still holds when $t-s$ is not divided by $\Delta B$, since it uses $\left\lfloor (t-s)/\Delta B \right\rfloor$ instead of $(t-s)/\Delta B$. Therefore, for all $t > s \geq 0$, we have
\begin{eqnarray}\label{lemma4/3}
\max_{1 \leq i \leq N} \textbf{x}_{i}(t) - \min_{1 \leq i \leq N} \textbf{x}_{i}(t) &\leq& \left(1 - \delta^{\Delta B}\right)^{\left\lfloor (t-s)/\Delta B \right\rfloor} \left(\max_{1 \leq i \leq N} \textbf{x}_{i}(s) - \min_{1 \leq i \leq N} \textbf{x}_{i}(s)\right) \nonumber \\
&\leq& 2 \left(\left(1 - \delta^{\Delta B}\right)^{\frac{1}{\Delta B}}\right)^{(t-s)} \left(\max_{1 \leq i \leq N} \textbf{x}_{i}(s) - \min_{1 \leq i \leq N} \textbf{x}_{i}(s)\right)
\end{eqnarray}

Equation (\ref{lemma4/3}) holds for any $\textbf{x}(s)$. By choosing $\textbf{x}(s)$ to be each of the $N$ basis vectors, (\ref{lemma4/3}) shows that for all $j \in \{1, 2, \ldots, N\}$,
\begin{equation}\label{lemma4/4}
\max_{1 \leq i \leq N} \left[(\textbf{W}^{(t)})^{T} \cdots (\textbf{W}^{(s)})^{T}\right]_{ij} - \min_{1 \leq i \leq N} \left[(\textbf{W}^{(t)})^{T} \cdots (\textbf{W}^{(s)})^{T}\right]_{ij} \leq 2 \left(\left(1 - \delta^{\Delta B}\right)^{\frac{1}{\Delta B}}\right)^{(t-s)}
\end{equation}

Since each $(\textbf{W}^{(i)})^{T}$ is row stochastic, let $\boldsymbol{\phi}_{j}(s)$ be the limit of the convex combinations of the $N$ numbers $\left[(\textbf{W}^{(t)})^{T} \cdots (\textbf{W}^{(s)})^{T}\right]_{ij}$, for $i \in \{1, \ldots, N\}$, as $t \to \infty$. Hence, we have proven the lemma for $t > s$. For $t = s$, it is obvious that $\left| (\textbf{W}^{(s)})^{T}_{ij} - \boldsymbol{\phi}_{j}(s) \right| \leq 2$, since $(\textbf{W}^{(s)})^{T}$ is row stochastic and $\boldsymbol{\phi}(s)$ is a stochastic vector, the lemma also holds.
\end{proof}

\noindent
By taking the transpose of the multiplication of the row stochastic matrix sequence in Lemma 7, we obtain the Lemma 1 about the column stochastic matrix sequence.

\section{Proof of Lemma 3}
\noindent
{\bf Corollary 4} {\it
Suppose that the graph sequence $\{\mathcal{G}^{(t)}\}$ satisfies Assumption 1, and the weighting matrix sequence $\{\textbf{W}^{(t)}\}$ is defined by Definition 1. For $t<\Delta B$, the distance between the node's corrected parameter and the network average parameter is bounded by:
\begin{eqnarray}\label{L2_dis_equ}
\left \| \textbf{y} _{i}^{(t)}-\frac{1}{N}\sum_{i=1}^{N}\textbf{x}^{(t)}_{i}\right \|_{2} \le C\sqrt{N}\lambda^{t-1}\left \|\textbf{X}^{(0)} \right \|_{F}+C\sqrt{N}\sum_{s=1}^{t}\lambda^{t-s}\left \|\varepsilon^{(s)}\right \|_{2},\nonumber
\end{eqnarray}

\noindent
For $t\ge \Delta B$:
\begin{eqnarray}\label{L2_dis_equ2}
\left \| \textbf{y} _{i}^{(t)}-\frac{1}{N}\sum_{i=1}^{N}\textbf{x}^{(t)}_{i}\right \|_{2} \le \frac{C}{\sqrt{N}}\lambda^{t-1}\left \|\textbf{X}^{(0)} \right \|_{F}+\frac{C}{\sqrt{N}}\sum_{s=1}^{t}\lambda^{t-s}\left \|\boldsymbol{\varepsilon}^{(s)}\right \|_{2},\nonumber
\end{eqnarray}
where $C=\frac{4}{\delta^{\Delta B}}$, $\lambda=(1-\delta^{\Delta B})^{\frac{1}{\Delta B}}\in (0,1)$.
}
\begin{proof}
Same with the proof of Theorem 1 in the paper, we have
\begin{equation}\label{L2_dis_proof_1}
\begin{aligned}
\left ( \textbf{y}_{i}^{(t)} \right ) ^T-\frac{\textbf{1}^{T} }{N}\textbf{X}^{(t)}&=\frac{N\left [ \textbf{D}[t:1] \right ]_{i}\textbf{X}^{(0)}+N\sum_{s=1}^{t}\left [ \textbf{D}[t:s] \right ]_{i}\boldsymbol{\varepsilon}^{(s)}}{N\left (\boldsymbol{\phi}_{i} (t)N+\left [ \textbf{D}[t:1]\textbf{1}\right ]_{i}\right )} \\
&\quad -\frac{\textbf{1}^{T}\textbf{X}^{(t)}\left [ \textbf{D}[t:1]\textbf{1}\right ]_{i}}{N\left (\boldsymbol{\phi}_{i} (t)N+\left [ \textbf{D}[t:1]\textbf{1}\right ]_{i}\right )}
\end{aligned}
\end{equation}

\noindent
Take the Euclidean norm to both sides of Eq.\eqref{L2_dis_proof_1}.

\noindent
For $t< \Delta B$

\begin{equation}\label{L2_dis_proof_2}
\begin{aligned}
&\left \| \textbf{y}_{i}^{(t)}-\frac{\textbf{1}^{T} }{N}\textbf{X}^{(t)} \right \| _{2} \\
&\overset{(a)}{\le }\frac{1}{\delta ^{\Delta B}}\left \| \left [ \textbf{D}[t:1] \right ]_{i}\textbf{X}^{(0)} \right \|_{2}+\frac{1}{\delta ^{\Delta B}}\sum_{s=1}^{t}\left \| \left [ \textbf{D}[t:s] \right ]_{i}\boldsymbol{\varepsilon}^{(s)}\right \|_{2} \\
&\quad +\frac{1}{N\delta ^{\Delta B}}\left \| \textbf{1}^{T}\textbf{X}^{(t)}\left [ \textbf{D}[t:1]\right ]_{i}\textbf{1} \right \|_{2} \\
&\overset{(b)}{\le }\frac{1}{\delta ^{\Delta B}}\left \| \left [ \textbf{D}[t:1] \right ]_{i}\right \|_{2}\left \|\textbf{X}^{(0)} \right \|_{F}+\frac{1}{\delta ^{\Delta B}}k\lambda^{t-1}\left \| \textbf{1}^{T}\textbf{X}^{(t)} \right \|_{2}\\
&\quad +\frac{1}{\delta ^{\Delta B}}\sum_{s=1}^{t}\left \| \left [ \textbf{D}[t:s] \right ]_{i}\right \|_{2}\left \|\boldsymbol{\varepsilon}^{(s)}\right \|_{F}  \\
&\le \frac{k\sqrt{N}}{\delta ^{\Delta B}}\lambda^{t-1}\left \|\textbf{X}^{(0)} \right \|_{F}+\frac{k\sqrt{N}}{\delta ^{\Delta B}}\sum_{s=1}^{t}\lambda^{t-s}\left \|\boldsymbol{\varepsilon}^{(s)}\right \|_{F} \\
&\quad +\frac{1}{\delta ^{\Delta B}}k\lambda^{t-1}\left \| \textbf{1}^{T}\textbf{X}^{(t)} \right \|_{2}
\end{aligned}
\end{equation}
(a): By $\boldsymbol{\phi}_{i}(t)N+\left [ \textbf{D}[t:1]\textbf{1}\right ]_{i}=\left [ \textbf{W}[t:1]\textbf{1}\right ]_{i}$ and Lemma 2.\par
\noindent
(b): For arbitrary vector $\textbf{a}$ and matrix $\textbf{A}$, $\left \| \textbf{a}\textbf{A}  \right \|_{2}\le \left \| \textbf{a} \right \|_{2} \left \| \textbf{A} \right \|_{F} $.

\noindent
For $\left \| \textbf{1}^{T}\textbf{X}^{(t)} \right \|_{2}$ in Eq.\eqref{L2_dis_proof_2},
\begin{eqnarray}\label{L2_dis_proof_3}
\left \| \textbf{1}^{T}\textbf{X}^{(t)} \right \|_{2}&\overset{(a)}{=}&\left \| \textbf{1}^{T}\textbf{X}^{(0)}+\sum_{s=1}^{t}\textbf{1}^{T}\boldsymbol{\varepsilon}^{(s)} \right \|_{2} \nonumber \\
&\overset{(b)}{\le }&\sqrt{N}\left \| \textbf{X}^{(0)} \right \|_{F}+\sqrt{N}\sum_{s=1}^{t}\left \| \boldsymbol{\varepsilon}^{(s)} \right \|_{F}    
\end{eqnarray}
(a): $\textbf{W}[t:s]$ is a column stochastic matrix.\par
\noindent
(b): For arbitrary vector $\textbf{a}$ and matrix $\textbf{A}$, $\left \| \textbf{a}\textbf{A}  \right \|_{2}\le \left \| \textbf{a} \right \|_{2} \left \| \textbf{A} \right \|_{F} $.

\noindent
Take Eq.\eqref{L2_dis_proof_3} into Eq.\eqref{L2_dis_proof_2}, we have
\begin{eqnarray}\label{L2_dis_proof_4}
&&\left \| \textbf{y}_{i}^{(t)}-\frac{\textbf{1}^{T} }{N}\textbf{X}^{(t)} \right \| _{2}\nonumber \\
&&\overset{(a)}{\le} \frac{2k\sqrt{N}}{\delta ^{\Delta B}}\lambda^{t-1}\left \|\textbf{X}^{(0)} \right \|_{F}+\frac{2k\sqrt{N}}{\delta ^{\Delta B}}\sum_{s=1}^{t}\lambda^{t-s}\left \|\boldsymbol{\varepsilon}^{(s)}\right \|_{F}\nonumber\\
&&\overset{(b)}{\le} C\sqrt{N}\lambda^{t-1}\left \|\textbf{X}^{(0)} \right \|_{F}+C\sqrt{N}\sum_{s=1}^{t}\lambda^{t-s}\left \|\boldsymbol{\varepsilon}^{(s)}\right \|_{F}
\end{eqnarray}
(a): $\lambda<1$.\par
\noindent
(b): Let $C:=\frac{2k}{\delta^{\Delta B}}=\frac{4}{\delta^{\Delta B}}$.\par

\noindent
For $t\ge \Delta B$, take Lemma 2 into Eq.(\ref{L2_dis_proof_2}), we have
\begin{equation}\label{L2_dis_proof_5}
\begin{aligned}
&\left \| \textbf{y}_{i}^{(t)}-\frac{\textbf{1}^{T} }{N}\textbf{X}^{(t)} \right \| _{2} \\
&\overset{(a)}{\le }\frac{\left \| \left [ \textbf{D}[t:1] \right ]_{i}\textbf{X}^{(0)} \right \|_{2}}{N\delta ^{\Delta B}}+\frac{1}{N\delta ^{\Delta B}}\sum_{s=1}^{t}\left \| \left [ \textbf{D}[t:s] \right ]_{i}\boldsymbol{\varepsilon}^{(s)}\right \|_{2} \\
&\quad +\frac{1}{N^{2}\delta ^{\Delta B}}\left \| \textbf{1}^{T}\textbf{X}^{(t)}\left [ \textbf{D}[t:1]\right ]_{i}\textbf{1} \right \|_{2}
\end{aligned}
\end{equation}
(a): By $\boldsymbol{\phi}_{i}(t)N+\left [ \textbf{D}[t:1]\textbf{1}\right ]_{i}=\left [ \textbf{W}[t:1]\textbf{1}\right ]_{i}$ and Lemma 2.\par

\noindent
In the same way as $t<\Delta B$, we can prove Corollary 4 in $t\ge \Delta B$.
\end{proof}

\noindent
Rewrite \texttt{MSGAP} to the matrix form, we have
\begin{equation}\label{ori_matrix}
\left\{\begin{matrix}
\textbf{M}^{(t)}=\beta \textbf{M}^{(t-1)}+\textbf{G}^{(t-1)}\\ 
\textbf{X}^{(t-\frac{1}{2})}=\textbf{X}^{(t-1)}-\gamma \textbf{M}^{(t)}\\
\textbf{a}^{(t)}=\textbf{W}^{(t)}\textbf{a}^{(t-1)}\\
\textbf{X}^{(t)}=\textbf{W}^{(t)}\textbf{X}^{(t-\frac{1}{2})}\\
\left [ \textbf{Y} \right ]_{i}^{(t)}= \left [ \textbf{X} \right ]_{i}^{(t)}/a_{i}^{(t)}
\end{matrix}\right.
\end{equation}

\noindent
where
\begin{equation*}\label{vector_matrix}
\begin{matrix}
\textbf{M}^{(t)}=\left [\textbf{m}_{1}^{(t)},\textbf{m}_{2}^{(t)},...,\textbf{m}_{N}^{(t)}\right ] ^{T},\textbf{G}^{(t)}=\left [  \textbf{g}_{1}^{(t)},\textbf{g}_{2}^{(t)},...,\textbf{g}_{N}^{(t)}\right ]^{T}
\end{matrix}
\end{equation*}

\noindent
{\bf Corollary 5} {\it
Suppose that the graph sequence $\{\mathcal{G}^{(t)}\}$ satisfies Assumption 1, the weighting matrix sequence $\{\textbf{W}^{(t)}\}$ is generated by $\{\mathcal{G}^{(t)}\}$. Then for \texttt{MSGAP} and $t < \Delta B$, the distance between the corrected parameter of node $i$ and the network average parameter is bounded by:

\begin{eqnarray}\label{coro_th3_equ}
\left \| \textbf{y} _{i}^{(t)}-\overline{\textbf{x}}^{(t)}\right \|_{2}&\le& C\sqrt{N}\lambda ^{t-1}\left \| \textbf{X}^{(0)} \right \|_{F} \nonumber\\
&+& \gamma C\sqrt{N}\sum_{s=0}^{t-1}(t-s)\alpha ^{t-s-1}\left \| \textbf{G}^{(s)} \right \|_{F}\nonumber
\end{eqnarray}

\noindent
For $t \ge \Delta B$:

\begin{eqnarray}\label{coro_th3_equ_lager}
\left \| \textbf{y} _{i}^{(t)}-\overline{\textbf{x}}^{(t)}\right \|_{2}&\le& C\frac{\lambda ^{t-1}}{\sqrt{N}}\left \| \textbf{X}^{(0)} \right \|_{F} \nonumber\\
&+& \gamma C\frac{1}{\sqrt{N}}\sum_{s=0}^{t-1}(t-s)\alpha ^{t-s-1}\left \| \textbf{G}^{(s)} \right \|_{F}\nonumber
\end{eqnarray}

\noindent
where $\alpha=max\left \{\lambda,\beta \right\}\in \left(0,1\right)$.
}

\begin{proof}
Following from \texttt{MSGAP}, replace perturbation $\boldsymbol{\varepsilon}^{(t)}$ in Corollary 4 with momentum term in \texttt{MSGAP}, we have
\begin{eqnarray}\label{coro_1_proof_equ_1}
\boldsymbol{\varepsilon}^{(t)}&=&-\gamma \textbf{M}^{(t)} \nonumber\\
&\overset{(a)}{=}& -\gamma\left (\beta^{t}\textbf{M}^{(0)}+\sum_{s=0}^{t-1}\beta^{t-s-1}\textbf{G}^{(s)}\right ) \nonumber\\
&\overset{(b)}{=}&-\gamma \sum_{s=0}^{t-1}\beta^{t-s-1}\textbf{G}^{(s)}
\end{eqnarray}
(a): Substitute $\textbf{M}^{(t)}$ until $\textbf{M}^{(0)}$ \\
(b): Follows from $\textbf{m}_{i}^{(t)}=\textbf{0}$ in Algorithm 2.

\noindent
Taking \eqref{coro_1_proof_equ_1} into Corollary 4, for $t<\Delta B$, we have
\begin{eqnarray}\label{coro_1_proof_equ_2}
&&\left \| \textbf{y}_{i}^{(t)}-\frac{\textbf{1}^{T} }{N}\textbf{X}^{(t)} \right \| _{2}\nonumber \\
&&\le C\sqrt{N}\lambda^{t-1}\left \|\textbf{X}^{(0)} \right \|_{F}+C\sqrt{N}\sum_{s=1}^{t}\lambda^{t-s}\gamma\sum_{l=0}^{s-1}\beta^{s-l-1}\left \|\textbf{G}^{(l)}\right \|_{F}\nonumber\\
&&\le C\sqrt{N}\lambda^{t-1}\left \|\textbf{X}^{(0)} \right \|_{F}+\gamma C\sqrt{N}\sum_{l=0}^{t-1}\left \|\textbf{G}^{(l)}\right \|_{F}\sum_{s=l+1}^{t}\lambda^{t-s}\beta^{s-l-1}\nonumber\\
&&\overset{(a)}{\le} C\sqrt{N}\lambda^{t-1}\left \|\textbf{X}^{(0)} \right \|_{F}+\gamma C\sqrt{N}\sum_{l=0}^{t-1}\left \|\textbf{G}^{(l)}\right \|_{F}\sum_{s=l+1}^{t}\alpha^{t-l-1}\nonumber\\
&&\le C\sqrt{N}\lambda^{t-1}\left \|\textbf{X}^{(0)} \right \|_{F}+\gamma C\sqrt{N}\sum_{s=0}^{t-1}(t-s)\alpha^{t-s-1}\left \|\textbf{G}^{(s)}\right \|_{F}.
\end{eqnarray}
(a): Lat $\alpha=max\left \{\lambda,\beta \right\}\in \left ( 0,1 \right )$

\noindent
For $t\geq \Delta B$, we can similarly prove.
\end{proof}

\noindent
{\bf Definition 5} {\it 
\begin{equation}\label{gra_mat_equ}
\left\{\begin{matrix}
\textbf{H}^{(t)}:=\left [ \triangledown f_{1}(\textbf{y}_{1}^{(t)}),\triangledown f_{2}(\textbf{y}_{2}^{(t)}),...,\triangledown f_{N}(\textbf{y}_{N}^{(t)}) \right]^{T}\\
\textbf{J}^{(t)}:=\left [ \triangledown f_{1}(\overline{\textbf{x}}^{(t)}),\triangledown f_{2}(\overline{\textbf{x}}^{(t)}),...,\triangledown f_{N}(\overline{\textbf{x}}^{(t)}) \right]^{T}
\end{matrix}\right.
\end{equation}
}

\noindent
For Definition 5, we have the following lemma.

\noindent
{\bf Lemma 9} {\it 
Under Assumption 2, 3, for the matrices defined in Definition 5, we have
\begin{equation}\label{the2_aux_equ}
\mathbb{E} \left [ \left \| \textbf{H}^{(t)}\right \|^{2}\right ]\leq8L^{2}\sum_{i=1}^{N}\mathbb{E} \left [ \left \| \textbf{y}_{i}^{(t)}-\overline{\textbf{x}}^{(t)} \right \|^{2}\right ]+4N\kappa ^{2}+4N\mathbb{E} \left [ \left \| \frac{1}{N}\sum_{i=1}^{N}\triangledown f_{i}(\textbf{y}_{i}^{(t)})\right \|^{2}\right ]
\end{equation}
}
\begin{proof}
\begin{eqnarray}\label{the2_aux_pro_1}
&&\mathbb{E} \left [ \left \| \textbf{H}^{(t)}\right \|^{2}\right ]\nonumber\\
&&\overset{(a)}{=}\mathbb{E} \left [ \left \| \textbf{H}^{(t)}-\textbf{J}^{(t)}+\textbf{J}^{(t)}-\textbf{U}\textbf{J}^{(t)}+\textbf{U}\textbf{J}^{(t)}-\textbf{U}\textbf{H}^{(t)}+\textbf{U}\textbf{H}^{(t)} \right \|^{2}\right ]\nonumber \\
&&\overset{(b)}{\le}4\mathbb{E} \left [ \left \|\textbf{H}^{(t)}-\textbf{J}^{(t)} \right \|^{2}\right ]+4\mathbb{E} \left [ \left \|\textbf{J}^{(t)}-\textbf{U}\textbf{J}^{(t)} \right \|^{2}\right ]+4\mathbb{E} \left [ \left \|\textbf{U}\textbf{J}^{(t)}-\textbf{U}\textbf{H}^{(t)} \right \|^{2}\right ]+4\mathbb{E} \left [ \left \|\textbf{U}\textbf{H}^{(t)} \right \|^{2}\right ]\nonumber \\
&&\overset{(c)}{\leq}8L^{2}\sum_{i=1}^{N}\mathbb{E} \left [ \left \| \textbf{y}_{i}^{(t)}-\overline{\textbf{x}}^{(t)} \right \|^{2}\right ]+4\mathbb{E} \left [ \left \|\textbf{J}^{(t)}-\textbf{U}\textbf{J}^{(t)} \right \|^{2}\right ]+4\mathbb{E} \left [ \left \|\textbf{U}\textbf{H}^{(t)} \right \|^{2}\right ]\nonumber \\
&&\overset{(d)}{\leq}8L^{2}\sum_{i=1}^{N}\mathbb{E} \left [ \left \| \textbf{y}_{i}^{(t)}-\overline{\textbf{x}}^{(t)} \right \|^{2}\right ]+4N\kappa^{2}+4\mathbb{E} \left [ \left \|\textbf{U}\textbf{H}^{(t)} \right \|^{2}\right ]\nonumber \\
&&\overset{(e)}{\leq}8L^{2}\sum_{i=1}^{N}\mathbb{E} \left [ \left \| \textbf{y}_{i}^{(t)}-\overline{\textbf{x}}^{(t)} \right \|^{2}\right ]+4N\kappa^{2}+4N\mathbb{E} \left [ \left \| \frac{1}{N}\sum_{i=1}^{N}\triangledown f_{i}(\textbf{y}_{i}^{(t)})\right \|^{2}\right ]
\end{eqnarray}
(a): $\textbf{U}$ is a matrix with all entries equal to $1/N$.\\
(b): By $\left \| \sum^{N}\textbf{a} \right \|^{2}\le N\sum^{N}\left \| \textbf{a} \right \|^{2}$, where $N=4$.\\
(c): By L-Smoothness in Assumption 2 and Fact 1.\\
(d): By Bounded variance in Assumption 3 and Fact 1. \\
(e): By Fact 1.
\end{proof}

\subsection{Main Proof of Lemma 3}
\begin{proof}
By Corollary 5, for $t < \Delta B$, we have
\begin{eqnarray}\label{big_lem1_pro_1}
&&\mathbb{E} \left[\left \| \textbf{y} _{i}^{(t)}-\overline{\textbf{x}}^{(t)}\right \|_{2}^{2}\right]\nonumber\\
&&\le \mathbb{E} \left[\left(C\sqrt{N}\lambda ^{t-1}\left \| \textbf{X}^{(0)} \right \|_{F}+\gamma C\sqrt{N}\sum_{s=0}^{t-1}(t-s)\alpha ^{t-s-1}\left \| \textbf{G}^{(s)} \right \|_{F}\right)^{2}\right]\nonumber \\
&&\le \mathbb{E} [( C\sqrt{N}\lambda ^{t-1}\left \| \textbf{X}^{(0)} \right \|_{F}+\gamma C\sqrt{N}\sum_{s=0}^{t-1}(t-s)\alpha ^{t-s-1}\left \| \textbf{G}^{(s)}-\textbf{H}^{(s)} \right \|_{F} \nonumber\\
&&\quad +\gamma C\sqrt{N}\sum_{s=0}^{t-1}(t-s)\alpha ^{t-s-1}\left \| \textbf{H}^{(s)} \right \|_{F})^{2}]
\end{eqnarray}

\noindent
By $\mathbb{E}\left[(\textbf{a}+\textbf{b}+\textbf{c})^{2}\right]\le 3\mathbb{E}\left[\textbf{a}^{2}\right]+3\mathbb{E}\left[\textbf{b}^{2}\right]+3\mathbb{E}\left[\textbf{c}^{2}\right]$, for \eqref{big_lem1_pro_1}, we have
\begin{eqnarray}\label{big_lem1_pro_2}
\mathbb{E}\left[\textbf{a}^{2}\right]&=&C^{2}N\lambda^{2(t-1)}\left \| \textbf{X}^{(0)} \right \|^{2}\nonumber \\
\mathbb{E}\left[\textbf{b}^{2}\right]&=&\gamma^{2}C^{2}N\mathbb{E}\left[\left( \sum_{s=0}^{t-1}(t-s)\alpha ^{t-s-1}\left \| \textbf{G}^{(s)}-\textbf{H}^{(s)} \right \| \right)^{2}\right]\nonumber \\
&=&\underbrace{\gamma^{2}C^{2}N \mathbb{E}\left[\sum_{s=0}^{t-1}(t-s)^{2}\alpha ^{2(t-s-1)}\left \| \textbf{G}^{(s)}-\textbf{H}^{(s)} \right \|^{2}\right]}_{\textbf{b}_{1}} \nonumber \\
&&+\underbrace{\gamma^{2}C^{2}N \mathbb{E}\left[\sum_{s=0}^{t-1}\sum_{l=0,l\ne s}^{t-1}(t-s)(t-l)\alpha ^{2t-s-l-2}\left \| \textbf{G}^{(s)}-\textbf{H}^{(s)} \right \|\left \| \textbf{G}^{(l)}-\textbf{H}^{(l)} \right \|\right]}_{\textbf{b}_{2}}\nonumber \\
\mathbb{E}\left[\textbf{c}^{2}\right]&=&\gamma^{2}C^{2}N \mathbb{E}\left[\left( \sum_{s=0}^{t-1}(t-s)\alpha ^{t-s-1}\left \| \textbf{H}^{(s)} \right \| \right)^{2}\right]\nonumber \\
&=&\underbrace{\gamma^{2}C^{2}N \mathbb{E}\left[\sum_{s=0}^{t-1}(t-s)^{2}\alpha ^{2(t-s-1)}\left \| \textbf{H}^{(s)} \right \|^{2}\right]}_{\textbf{c}_{1}} \nonumber \\
&&+\underbrace{\gamma^{2}C^{2}N \mathbb{E}\left[\sum_{s=0}^{t-1}\sum_{l=0,l\ne s}^{t-1}(t-s)(t-l)\alpha ^{2t-s-l-2}\left \| \textbf{H}^{(s)} \right \|\left \| \textbf{H}^{(l)} \right \|\right]}_{\textbf{c}_{2}}
\end{eqnarray}

\noindent
For every equality in \eqref{big_lem1_pro_2},we have
\begin{eqnarray}\label{big_lem1_pro_3}
\textbf{b}_{1}&=&\gamma^{2}C^{2}N \sum_{s=0}^{t-1}(t-s)^{2}\alpha ^{2(t-s-1)}\mathbb{E}\left[\left \| \textbf{G}^{(s)}-\textbf{H}^{(s)} \right \|^{2}\right]\nonumber\\
&\overset{(a)}{\le}&\gamma^{2}C^{2}N^{2}\sigma^{2}\sum_{s=0}^{t-1}(t-s)^{2}\alpha ^{2(t-s-1)}\nonumber\\
&\overset{(b)}{\le}&\frac{\gamma^{2}C^{2}N^{2}\sigma^{2}}{(1-\alpha)^{4}}
\end{eqnarray}
(a): By Fact 1 and Assumption 3.\\
(b): $\sum_{s=0}^{\infty }s^{2}\alpha^{2(s-1)}=\frac{\alpha^{2}+1}{\left ( 1-\alpha^{2} \right )^{3}}\le\frac{1}{(1-\alpha)^{4}}$, for $\forall \alpha\in (0,1)$.

\begin{eqnarray}\label{big_lem1_pro_4}
\textbf{b}_{2}&=&2\gamma^{2}C^{2}N \mathbb{E}\left[\sum_{s=0}^{t-1}\sum_{l=s+1}^{t-1}(t-s)(t-l)\alpha ^{2t-s-l-2}\left \| \textbf{G}^{(s)}-\textbf{H}^{(s)} \right \|\left \| \textbf{G}^{(l)}-\textbf{H}^{(l)} \right \| \right]\nonumber\\
&\overset{(a)}{\le}&\gamma^{2}C^{2}N \sum_{s=0}^{t-1}\sum_{l=s+1}^{t-1}(t-s)(t-l)\alpha ^{2t-s-l-2}\mathbb{E}\left[\left(\left \| \textbf{G}^{(s)}-\textbf{H}^{(s)} \right \|^{2}+\left \| \textbf{G}^{(l)}-\textbf{H}^{(l)} \right \|^{2}\right )\right]\nonumber\\
&\overset{(b)}{\le}&2\gamma^{2}C^{2}N^{2}\sigma^{2} \sum_{s=0}^{t-1}\sum_{l=s+1}^{t-1}(t-s)(t-l)\alpha ^{2t-s-l-2}\nonumber\\
&\overset{(c)}{\le}&\frac{2\gamma^{2}C^{2}N^{2}\sigma^{2}}{(1-\alpha)^{4}}
\end{eqnarray}
(a): $2\textbf{a}\textbf{b}\le \textbf{a}^{2}+\textbf{b}^{2}$\\
(b): By Fact 1 and Assumption 3\\
(c): $\sum_{n=0}^{\infty}n\alpha^{n-1}=\frac{1}{(1-\alpha)^{2}}$

\begin{eqnarray}\label{big_lem1_pro_5}
\textbf{c}_{1}&=&\gamma^{2}C^{2}N \sum_{s=0}^{t-1}(t-s)^{2}\alpha ^{2(t-s-1)}\mathbb{E}\left[\left \| \textbf{H}^{(s)} \right \|^{2}\right] \nonumber\\
&\overset{(a)}{\le}&\frac{\gamma^{2}C^{2}N}{(1-\alpha)^{2}}\sum_{s=0}^{t-1}(t-s)\alpha ^{t-s-1}\mathbb{E}\left[\left \| \textbf{H}^{(s)} \right \|^{2}\right]
\end{eqnarray}
(a):$(t-s)\alpha^{t-s-1}\le\sum_{(t-s)=0}^{\infty}(t-s)\alpha^{t-s-1}=\frac{1}{(1-\alpha)^{2}}$

\begin{eqnarray}\label{big_lem1_pro_6}
\textbf{c}_{2}&=&2\gamma^{2}C^{2}N \sum_{s=0}^{t-1}\sum_{l=s+1}^{t-1}(t-s)(t-l)\alpha ^{2t-s-l-2} \mathbb{E}\left[\left \| \textbf{H}^{(s)} \right \|\left \| \textbf{H}^{(l)} \right \|\right] \nonumber\\
&\le&\gamma^{2}C^{2}N \sum_{s=0}^{t-1}\sum_{l=s+1}^{t-1}(t-s)(t-l)\alpha ^{2t-s-l-2}\mathbb{E}\left[\left(\left \| \textbf{H}^{(s)} \right \|^{2}+\left \| \textbf{H}^{(l)} \right \|^{2}\right)\right] \nonumber\\
&\le&\gamma^{2}C^{2}N \sum_{s=0}^{t-1}\sum_{l=s+1}^{t-1}(t-s)(t-l)\alpha ^{2t-s-l-2}\mathbb{E}\left[\left \| \textbf{H}^{(s)} \right \|^{2}\right]\nonumber\\
&&+\gamma^{2}C^{2}N \sum_{s=0}^{t-1}\sum_{l=s+1}^{t-1}(t-s)(t-l)\alpha ^{2t-s-l-2}\mathbb{E}\left[\left \| \textbf{H}^{(l)} \right \|^{2}\right]\nonumber\\
&\le&2\gamma^{2}C^{2}N \sum_{s=0}^{t-1}\sum_{l=s+1}^{t-1}(t-s)(t-l)\alpha ^{2t-s-l-2}\mathbb{E}\left[\left \| \textbf{H}^{(s)} \right \|^{2}\right]\nonumber\\
&\overset{(a)}{\le}&\frac{2\gamma^{2}C^{2}N}{(1-\alpha)^{2}} \sum_{s=0}^{t-1}(t-s)\alpha ^{t-s-1}\mathbb{E}\left[\left \| \textbf{H}^{(s)} \right \|^{2}\right]
\end{eqnarray}
(a):$\sum(t-s)\alpha^{t-s-1}\le\sum_{(t-s)=0}^{\infty}(t-s)\alpha^{t-s-1}=\frac{1}{(1-\alpha)^{2}}$

\noindent
By $\mathbb{E}\left[(\textbf{a}+\textbf{b}+\textbf{c})^{2}\right]\le 3\mathbb{E}\left[\textbf{a}^{2}\right]+3\mathbb{E}\left[\textbf{b}^{2}\right]+3\mathbb{E}\left[\textbf{c}^{2}\right]$, for \eqref{big_lem1_pro_1}, we have
\begin{eqnarray}\label{big_lem1_pro_8}
\mathbb{E}\left [ \left \| \textbf{y}_{i}^{(t)}-\overline{\textbf{x}}^{(t)}\right \|^{2}\right ]&\le&  3C^{2}N\lambda^{2(t-1)} \left \| \textbf{X}^{(0)} \right \|^{2}+9\frac{\gamma^{2} C^{2}N^{2}}{(1-\alpha)^{4}}\sigma^{2}\nonumber \\
&&+9\frac{\gamma^{2} C^{2}N}{(1-\alpha)^{2}}\sum_{s=0}^{t-1}(t-s)\alpha^{t-s-1}\mathbb{E}\left[\left \| \textbf{H}^{(s)} \right \| ^{2}\right]
\end{eqnarray}

\noindent
Sum up \eqref{big_lem1_pro_8} over $i\in\{1,...,N\}$, we have the following inequality for $t<\Delta B$
\begin{eqnarray}\label{big_lem1_pro_9}
\sum_{i=1}^{N}\mathbb{E}\left [ \left \| \textbf{y}_{i}^{(t)}-\overline{\textbf{x}}^{(t)}\right \|^{2}\right ]&\le& 3C^{2}N^{2}\lambda^{2(t-1)} \left \| \textbf{X}^{(0)} \right \|^{2}+9\frac{\gamma^{2} C^{2}N^{3}}{(1-\alpha)^{4}}\sigma^{2}\nonumber \\
&&+9\frac{\gamma^{2} C^{2}N^{2}}{(1-\alpha)^{2}}\sum_{s=0}^{t-1}(t-s)\alpha^{t-s-1}\mathbb{E}\left[\left \| \textbf{H}^{(s)} \right \| ^{2}\right]
\end{eqnarray}

\noindent
For $t\ge \Delta B$, by Corollary 5, in the same way of $t<\Delta B$, we have
\begin{eqnarray}\label{big_lem1_pro_10}
\sum_{i=1}^{N}\mathbb{E}\left [ \left \| \textbf{y}_{i}^{(t)}-\overline{\textbf{x}}^{(t)}\right \|^{2}\right ]&\le& 3C^{2}\lambda^{2(t-1)} \left \| \textbf{X}^{(0)} \right \|^{2}+9\frac{\gamma^{2} C^{2}N}{(1-\alpha)^{4}}\sigma^{2}\nonumber \\
&&+9\frac{\gamma^{2} C^{2}}{(1-\alpha)^{2}}\sum_{s=0}^{t-1}(t-s)\alpha^{t-s-1}\mathbb{E}\left[\left \| \textbf{H}^{(s)} \right \| ^{2}\right]
\end{eqnarray}

\noindent
Since $T>\Delta B$, sum up \eqref{big_lem1_pro_9} over $t\in\{0,...,T-1\}$, we have

\begin{eqnarray}\label{big_lem1_pro_11}
\sum_{t=0}^{T-1}\sum_{i=1}^{N}\mathbb{E}\left [ \left \| \textbf{y}_{i}^{(t)}-\overline{\textbf{x}}^{(t)}\right \|^{2}\right ]&=& \underbrace{\sum_{i=1}^{N}\mathbb{E}\left [ \left \| \textbf{y}_{i}^{(0)}-\overline{\textbf{x}}^{(0)}\right \|^{2}\right ]}_{\textbf{a}}+\underbrace{\sum_{t=1}^{\Delta B-1}\sum_{i=1}^{N}\mathbb{E}\left [ \left \| \textbf{y}_{i}^{(t)}-\overline{\textbf{x}}^{(t)}\right \|^{2}\right ]}_{\textbf{b}}\nonumber\\
&&+\underbrace{\sum_{t=\Delta B}^{T-1}\sum_{i=1}^{N}\mathbb{E}\left [ \left \| \textbf{y}_{i}^{(t)}-\overline{\textbf{x}}^{(t)}\right \|^{2}\right ]}_{\textbf{c}}
\end{eqnarray}

\noindent
For \textbf{a} in \eqref{big_lem1_pro_11}
\begin{eqnarray}\label{big_lem1_pro_12}
\sum_{i=1}^{N}\mathbb{E}\left [ \left \| \textbf{y}_{i}^{(0)}-\overline{\textbf{x}}^{(0)}\right \|^{2}\right ] &=& \sum_{i=1}^{N} \left \| \textbf{y}_{i}^{(0)}-\overline{\textbf{x}}^{(0)}\right \|^{2}\nonumber\\
&\overset{(a)}{=}& \left \| \textbf{Y}^{(0)}-\overline{\textbf{X}}^{(0)}   \right \| ^{2}\nonumber\\
&\overset{(b)}{=}& \left \|\left ( \textbf{I}-\textbf{U}   \right )  \textbf{X}^{(0)}   \right \| ^{2}\nonumber\\
&\overset{(c)}{\le}& \left \| \textbf{I}-\textbf{U} \right \| ^{2}\left \| \textbf{X}^{(0)} \right \| ^{2} \nonumber\\
&\le &2N\left \| \textbf{X}^{(0)} \right \| ^{2}
\end{eqnarray}
(a): By Fact 1.\\
\noindent
(b): $\textbf{I}$ is the identity matrix, $\textbf{U}$ is a matrix with all entries equal to $1/N$.\\
\noindent
(c): The sub-multiplicative property of Frobenius norm.

\noindent
For \textbf{b} in \eqref{big_lem1_pro_11}, by \eqref{big_lem1_pro_9}, we have
\begin{eqnarray}\label{big_lem1_pro_13}
\sum_{t=1}^{\Delta B-1}\sum_{i=1}^{N}\mathbb{E}\left [ \left \| \textbf{y}_{i}^{(t)}-\overline{\textbf{x}}^{(t)}\right \|^{2}\right ] &\overset{(a)}{\le}& \frac{3C^{2}N^{2}}{1-\lambda^{2}} \left \| \textbf{X}^{(0)} \right \|^{2} + 9\frac{\gamma^{2} C^{2}N^{3}\Delta B}{(1-\alpha)^{4}}\sigma^{2}\nonumber\\
&& +9\frac{\gamma^{2} C^{2}N^{2}}{(1-\alpha)^{2}}\underbrace{\sum_{t=1}^{\Delta B}\sum_{s=0}^{t-1}(t-s)\alpha^{t-s-1}\mathbb{E}\left[\left \| \textbf{H}^{(s)} \right \| ^{2}\right]}_{\textbf{b}_{1}}
\end{eqnarray}
(a): $\sum_{s=0}^{\infty }\lambda^{2s}=\frac{1}{1-\lambda^{2}}$

\noindent
For $\textbf{b}_{1}$ in \eqref{big_lem1_pro_13}
\begin{eqnarray}\label{big_lem1_pro_14}
\textbf{b}_{1}&\le& \sum_{t=1}^{\Delta B}\sum_{s=0}^{t-1}(t-s)\alpha^{t-s-1}\mathbb{E}\left[\left \| \textbf{H}^{(s)} \right \| ^{2}\right] \nonumber\\
&\le& \sum_{s=0}^{\Delta B-1} \mathbb{E}\left[\left \| \textbf{H}^{(s)} \right \| ^{2}\right] \sum_{t=s+1}^{\Delta B}(t-s)\alpha^{t-s-1}\nonumber\\
&\overset{(a)}{\le}& \frac{1}{(1-\alpha)^{2}}\sum_{t=0}^{\Delta B-1} \mathbb{E}\left[\left \| \textbf{H}^{(t)} \right \| ^{2}\right] 
\end{eqnarray}
(a):$\sum(t-s)\alpha^{t-s-1}\le\sum_{(t-s)=0}^{\infty}(t-s)\alpha^{t-s-1}=\frac{1}{(1-\alpha)^{2}}$

\noindent
By Lemma 9, we have
\begin{eqnarray}\label{big_lem1_pro_15}
\textbf{b}_{1}&\le& \frac{8L^{2}}{(1-\alpha)^{2}}\sum_{t=0}^{\Delta B-1}\sum_{i=1}^{N}\mathbb{E}\left [ \left \| \textbf{y}_{i}^{(t)}-\overline{\textbf{x}}^{(t)}\right \|^{2}\right ]+\frac{4\Delta B N}{(1-\alpha)^{2}} \kappa ^{2}\nonumber\\
&&+\frac{4N}{(1-\alpha)^{2}} \sum_{t=0}^{\Delta B-1}\mathbb{E}\left [ \left \| \frac{1}{N}\sum_{i=1}^{N}\triangledown f_{i}(\textbf{y}_{i}^{(t)})\right \|^{2}\right ]\nonumber \\
&\le& \frac{8L^{2}}{(1-\alpha)^{2}}\sum_{t=0}^{T-1}\sum_{i=1}^{N}\mathbb{E}\left [ \left \| \textbf{y}_{i}^{(t)}-\overline{\textbf{x}}^{(t)}\right \|^{2}\right ]+\frac{4\Delta B N}{(1-\alpha)^{2}} \kappa ^{2}\nonumber\\
&&+\frac{4N}{(1-\alpha)^{2}} \sum_{t=0}^{T-1}\mathbb{E}\left [ \left \| \frac{1}{N}\sum_{i=1}^{N}\triangledown f_{i}(\textbf{y}_{i}^{(t)})\right \|^{2}\right ]
\end{eqnarray}

\noindent
Take \eqref{big_lem1_pro_15} into \eqref{big_lem1_pro_13}, we have the upper bound of \textbf{b} in \eqref{big_lem1_pro_11}
\begin{eqnarray}\label{big_lem1_pro_16}
\sum_{t=1}^{\Delta B-1}\sum_{i=1}^{N}\mathbb{E}\left [ \left \| \textbf{y}_{i}^{(t)}-\overline{\textbf{x}}^{(t)}\right \|^{2}\right ]&\le& \frac{3C^{2}N^{2}}{1-\lambda^{2}} \left \| \textbf{X}^{(0)} \right \|^{2} + 9\frac{\gamma^{2} C^{2}N^{3}\Delta B}{(1-\alpha)^{4}}\sigma^{2}+36\frac{\gamma^{2} C^{2}N^{3}\Delta B}{(1-\alpha)^{4}}\kappa ^{2}\nonumber\\
&&+72\frac{\gamma^{2} C^{2}N^{2}L^{2}}{(1-\alpha)^{4}}\sum_{t=0}^{T-1}\sum_{i=1}^{N}\mathbb{E}\left [ \left \| \textbf{y}_{i}^{(t)}-\overline{\textbf{x}}^{(t)}\right \|^{2}\right ]\nonumber\\
&&+36\frac{\gamma^{2} C^{2}N^{3}}{(1-\alpha)^{4}}\sum_{t=0}^{T-1}\mathbb{E}\left [ \left \| \frac{1}{N}\sum_{i=1}^{N}\triangledown f_{i}(\textbf{y}_{i}^{(t)})\right \|^{2}\right ]
\end{eqnarray}

\noindent
For \textbf{c} in \eqref{big_lem1_pro_11}, by \eqref{big_lem1_pro_10}, we have
\begin{eqnarray}\label{big_lem1_pro_17}
\sum_{t=\Delta B}^{T-1}\sum_{i=1}^{N}\mathbb{E}\left [ \left \| \textbf{y}_{i}^{(t)}-\overline{\textbf{x}}^{(t)}\right \|^{2}\right ] &\overset{(a)}{\le}& \frac{3C^{2}}{1-\lambda^{2}} \left \| \textbf{X}^{(0)} \right \|^{2} + 9\frac{\gamma^{2} C^{2}NT}{(1-\alpha)^{4}}\sigma^{2}\nonumber\\
&& +9\frac{\gamma^{2} C^{2}}{(1-\alpha)^{2}}\underbrace{\sum_{t=\Delta B}^{T-1}\sum_{s=0}^{t-1}(t-s)\alpha^{t-s-1}\mathbb{E}\left[\left \| \textbf{H}^{(s)} \right \| ^{2}\right]}_{\textbf{c}_{1}}
\end{eqnarray}
(a): $\sum_{s=0}^{\infty }\lambda^{2s}=\frac{1}{1-\lambda^{2}}$

\noindent
For $\textbf{c}_{1}$ in \eqref{big_lem1_pro_17}
\begin{eqnarray}\label{big_lem1_pro_18}
\textbf{c}_{1}&\le& \sum_{t=1}^{T}\sum_{s=0}^{t-1}(t-s)\alpha^{t-s-1}\mathbb{E}\left[\left \| \textbf{H}^{(s)} \right \| ^{2}\right] \nonumber\\
&\le& \sum_{s=0}^{T-1} \mathbb{E}\left[\left \| \textbf{H}^{(s)} \right \| ^{2}\right] \sum_{t=s+1}^{T}(t-s)\alpha^{t-s-1}\nonumber\\
&\overset{(a)}{\le}& \frac{1}{(1-\alpha)^{2}}\sum_{t=0}^{T-1} \mathbb{E}\left[\left \| \textbf{H}^{(t)} \right \| ^{2}\right] 
\end{eqnarray}
(a):$\sum(t-s)\alpha^{t-s-1}\le\sum_{(t-s)=0}^{\infty}(t-s)\alpha^{t-s-1}=\frac{1}{(1-\alpha)^{2}}$

\noindent
By Lemma 9, we have
\begin{eqnarray}\label{big_lem1_pro_19}
\textbf{c}_{1}&\le& \frac{8L^{2}}{(1-\alpha)^{2}}\sum_{t=0}^{T-1}\sum_{i=1}^{N}\mathbb{E}\left [ \left \| \textbf{y}_{i}^{(t)}-\overline{\textbf{x}}^{(t)}\right \|^{2}\right ]+\frac{4 NT}{(1-\alpha)^{2}} \kappa ^{2}\nonumber\\
&&+\frac{4N}{(1-\alpha)^{2}} \sum_{t=0}^{T-1}\mathbb{E}\left [ \left \| \frac{1}{N}\sum_{i=1}^{N}\triangledown f_{i}(\textbf{y}_{i}^{(t)})\right \|^{2}\right ]
\end{eqnarray}

\noindent
Take \eqref{big_lem1_pro_19} into \eqref{big_lem1_pro_17}, we have the upper bound of \textbf{c} in \eqref{big_lem1_pro_11}
\begin{eqnarray}\label{big_lem1_pro_20}
\sum_{t=\Delta B}^{T-1}\sum_{i=1}^{N}\mathbb{E}\left [ \left \| \textbf{y}_{i}^{(t)}-\overline{\textbf{x}}^{(t)}\right \|^{2}\right ]&\le& \frac{3C^{2}}{1-\lambda^{2}} \left \| \textbf{X}^{(0)} \right \|^{2} + 9\frac{\gamma^{2} C^{2}NT}{(1-\alpha)^{4}}\sigma^{2}+36\frac{\gamma^{2} C^{2}NT}{(1-\alpha)^{4}}\kappa ^{2}\nonumber\\
&&+72\frac{\gamma^{2} C^{2}L^{2}}{(1-\alpha)^{4}}\sum_{t=0}^{T-1}\sum_{i=1}^{N}\mathbb{E}\left [ \left \| \textbf{y}_{i}^{(t)}-\overline{\textbf{x}}^{(t)}\right \|^{2}\right ]\nonumber\\
&&+36\frac{\gamma^{2} C^{2}N}{(1-\alpha)^{4}}\sum_{t=0}^{T-1}\mathbb{E}\left [ \left \| \frac{1}{N}\sum_{i=1}^{N}\triangledown f_{i}(\textbf{y}_{i}^{(t)})\right \|^{2}\right ]\nonumber \\
&\le& \frac{3C^{2}N^{2}}{1-\lambda^{2}} \left \| \textbf{X}^{(0)} \right \|^{2} + 9\frac{\gamma^{2} C^{2}NT}{(1-\alpha)^{4}}\sigma^{2}+36\frac{\gamma^{2} C^{2}NT}{(1-\alpha)^{4}}\kappa ^{2}\nonumber\\
&&+72\frac{\gamma^{2} C^{2}N^{2}L^{2}}{(1-\alpha)^{4}}\sum_{t=0}^{T-1}\sum_{i=1}^{N}\mathbb{E}\left [ \left \| \textbf{y}_{i}^{(t)}-\overline{\textbf{x}}^{(t)}\right \|^{2}\right ]\nonumber\\
&&+36\frac{\gamma^{2} C^{2}N^{3}}{(1-\alpha)^{4}}\sum_{t=0}^{T-1}\mathbb{E}\left [ \left \| \frac{1}{N}\sum_{i=1}^{N}\triangledown f_{i}(\textbf{y}_{i}^{(t)})\right \|^{2}\right ]
\end{eqnarray} 

\noindent
Take \textbf{a}, \textbf{b}, \textbf{c}, into \eqref{big_lem1_pro_11}, we have
\begin{eqnarray}\label{big_lem1_pro_21}
&&\sum_{t=0}^{T-1}\sum_{i=1}^{N}\mathbb{E}\left [ \left \| \textbf{y}_{i}^{(t)}-\overline{\textbf{x}}^{(t)}\right \|^{2}\right ]\nonumber \\
&&\le \left ( 2N+ \frac{6C^{2}N^{2}}{1-\lambda^{2}}\right )   \left \| \textbf{X}^{(0)} \right \|^{2}+144\frac{\gamma^{2} C^{2}N^{2}L^{2}}{(1-\alpha)^{4}}\sum_{t=0}^{T-1}\sum_{i=1}^{N}\mathbb{E}\left [ \left \| \textbf{y}_{i}^{(t)}-\overline{\textbf{x}}^{(t)}\right \|^{2}\right ]\nonumber\\
&&\quad +72\frac{\gamma^{2} C^{2}N^{3}}{(1-\alpha)^{4}}\sum_{t=0}^{T-1}\mathbb{E}\left [ \left \| \frac{1}{N}\sum_{i=1}^{N}\triangledown f_{i}(\textbf{y}_{i}^{(t)})\right \|^{2}\right ]\nonumber \\
&&\quad +9\frac{\gamma^{2} C^{2}N}{(1-\alpha)^{4}}\left ( T+N^{2}\Delta B \right ) \sigma^{2}+36\frac{\gamma^{2} C^{2}N}{(1-\alpha)^{4}}\left ( T+N^{2}\Delta B \right )\kappa ^{2}
\end{eqnarray} 

\noindent
Rearranging \eqref{big_lem1_pro_21}, we have
\begin{eqnarray}\label{big_lem1_pro_22}
&&Q\sum_{t=0}^{T-1}\sum_{i=1}^{N}\mathbb{E}\left [ \left \| \textbf{y}_{i}^{(t)}-\overline{\textbf{x}}^{(t)}\right \|^{2}\right ]\nonumber\\
&&\le\left(2N+\frac{6C^{2}N^{2}}{1-\lambda ^{2}}\right) \left \| \textbf{X}^{(0)} \right \|^{2}+72\frac{\gamma^{2} C^{2}N^{3}}{(1-\alpha)^{4}}\sum_{t=0}^{T-1}\mathbb{E}\left [ \left \| \frac{1}{N}\sum_{i=1}^{N}\triangledown f_{i}(\textbf{y}_{i}^{(t)})\right \|^{2}\right ]\nonumber\\
&&\quad +9\frac{\gamma^{2} C^{2}N}{(1-\alpha)^{4}}\left ( T+N^{2}\Delta B \right ) \sigma^{2}+36\frac{\gamma^{2} C^{2}N}{(1-\alpha)^{4}}\left ( T+N^{2}\Delta B \right )\kappa ^{2}
\end{eqnarray}
where $Q:=1-144\frac{\gamma^{2} C^{2}N^{2}L^{2}}{(1-\alpha)^{4}}$, when $\gamma<\frac{(1-\alpha)^{2}}{12CNL}$, we have $Q>0$.\par

\noindent
Divide $Q$ on both sides of \eqref{big_lem1_pro_22} with $\gamma<\frac{(1-\alpha)^{2}}{12CNL}$, we have
\begin{eqnarray}\label{big_lem1_pro_23}
&&\sum_{t=0}^{T-1}\sum_{i=1}^{N}\mathbb{E}\left [ \left \| \textbf{y}_{i}^{(t)}-\overline{\textbf{x}}^{(t)}\right \|^{2}\right ]\nonumber\\
&&\le\frac{1}{Q}\left(2N+\frac{6C^{2}N^{2}}{1-\lambda ^{2}}\right) \left \| \textbf{X}^{(0)} \right \|^{2}+72\frac{\gamma^{2} C^{2}N^{3}}{Q(1-\alpha)^{4}}\sum_{t=0}^{T-1}\mathbb{E}\left [ \left \| \frac{1}{N}\sum_{i=1}^{N}\triangledown f_{i}(\textbf{y}_{i}^{(t)})\right \|^{2}\right ]\nonumber\\
&&\quad +9\frac{\gamma^{2} C^{2}N}{Q(1-\alpha)^{4}}\left ( T+N^{2}\Delta B \right ) \sigma^{2}+36\frac{\gamma^{2} C^{2}N}{Q(1-\alpha)^{4}}\left ( T+N^{2}\Delta B \right )\kappa ^{2}
\end{eqnarray}
Here, we complete the proof.
\end{proof}

\section{Proof of Lemma 4}
\begin{proof}
\begin{eqnarray}\label{op1_the_pro_10}
&&\mathbb{E}\left [  \left \| \frac{1}{N}\sum_{i=1}^{N}\textbf{g}_{i}^{(t)} \right \|^{2}\right ]\nonumber \\
&&\overset{(a)}{=} \mathbb{E}\left [  \left \| \frac{1}{N}\sum_{i=1}^{N}\left ( \textbf{g}_{i}^{(t)}-\triangledown f_{i}(\textbf{y}_{i}^{(t)} ) \right )  \right \|^{2}\right ]+\mathbb{E}\left [  \left \| \frac{1}{N}\sum_{i=1}^{N}\triangledown f_{i}(\textbf{y}_{i}^{(t)}) \right \|^{2}\right ]\nonumber \\
&&\overset{(b)}{=}\frac{1}{N^{2}}\sum_{i=1}^{N}\mathbb{E}\left [ \left \|  \textbf{g}_{i}^{(t)}-\triangledown f_{i}(\textbf{y}_{i}^{(t)} ) \right \|^{2} \right ]+\mathbb{E}\left [  \left \| \frac{1}{N}\sum_{i=1}^{N}\triangledown f_{i}(\textbf{y}_{i}^{(t)}) \right \|^{2}\right ]\nonumber \\
&&\overset{(c)}{\leq}\frac{1}{N}\sigma^{2}+\mathbb{E}\left [  \left \| \frac{1}{N}\sum_{i=1}^{N}\triangledown f_{i}(\textbf{y}_{i}^{(t)}) \right \|^{2}\right ]
\end{eqnarray}
(a): Follows by $\mathbb{E}\left [ \textbf{g}_{i}  \right ] =\triangledown f_{i}(\textbf{y}_{i} )$.\\
(b): Follows by $\mathbb{E}\left [ \textbf{g}_{i}  \right ] =\triangledown f_{i}(\textbf{y}_{i} )$.\\
(c): Follows by Assumption 3.
\end{proof}

\section{Proof of Lemma 5}
\noindent
{\bf Lemma 10} {\it 
Consider $\overline{\textbf{m}}^{(t)},\overline{\textbf{x}}^{(t)}$ in \texttt{MSGAP}, for all $t\geq 1$, we have
\begin{eqnarray}\label{lem_ave_equ}
\left\{\begin{array}{lr}
\overline{\textbf{m}}^{(t)}=\beta \overline{\textbf{m}}^{(t-1)}+\frac{1}{N}\sum_{i=1}^{N}\textbf{g}_{i}^{(t-1)}\\ 
\overline{\textbf{x}}^{(t)}=\overline{\textbf{x}}^{(t-1)}-\gamma \overline{\textbf{m}}^{(t)}
\end{array}\right.\nonumber
\end{eqnarray}
}

\begin{proof} From \texttt{MSGAP}, we have 
\begin{eqnarray}\label{lem_ave_proof_1}
\textbf{m}_{i}^{(t)}=\beta \textbf{m}_{i}^{(t-1)}+\textbf{g}_{i}^{(t-1)}
\end{eqnarray}

\noindent
Summing \eqref{lem_ave_proof_1} over $i\in \left \{ 1,2,...,N \right \} $, we have
\begin{eqnarray}\label{lem_ave_proof_2}
\sum_{i=1}^{N}\textbf{m}_{i}^{(t)}=\beta \sum_{i=1}^{N}\textbf{m}_{i}^{(t-1)}+\sum_{i=1}^{N}\textbf{g}_{i}^{(t-1)}
\end{eqnarray}

\noindent
Dividing both sides by $N$, we have
\begin{eqnarray}\label{lem_ave_proof_3}
\frac{1}{N}\sum_{i=1}^{N}\textbf{m}_{i}^{(t)}=\beta \frac{1}{N}\sum_{i=1}^{N}\textbf{m}_{i}^{(t-1)}+\frac{1}{N}\sum_{i=1}^{N}\textbf{g}_{i}^{(t-1)}
\end{eqnarray}

\noindent
By the definition of $\overline{\textbf{m}}^{(t)}$, we have the first part of Lemma 10. As for the second part of Lemma 10, from the definition of $\overline{\textbf{x}}^{(t)}$, we have
\begin{eqnarray}\label{lem_ave_proof_4}
\overline{\textbf{x}}^{(t)}&=&\frac{1}{N}\sum_{i=1}^{N}\textbf{x}_{i}^{(t)} \nonumber \\
~&\overset{(25)}{=}&\frac{1}{N}\sum_{i=1}^{N}\sum_{j=1}^{N}w_{ij}^{(t)}\textbf{x}_{j}^{(t-\frac{1}{2})} \nonumber \\
~&=&\frac{1}{N} \sum_{j=1}^{N}\textbf{x}_{j}^{(t-\frac{1}{2})}\sum_{i=1}^{N}w_{ij}^{(t)} \nonumber \\
~&\overset{(a)}{=}&\frac{1}{N}\sum_{j=1}^{N}\textbf{x}_{j}^{(t-\frac{1}{2})} \nonumber \\
~&\overset{(23)}{=}&\frac{1}{N}\sum_{i=1}^{N}\left(\textbf{x}_{i}^{(t-1)}-\gamma \textbf{m}_{i}^{(t)}\right) \nonumber\\
~&=&\overline{\textbf{x}}^{(t-1)}-\gamma \overline{\textbf{m}}^{(t)}
\end{eqnarray}
(a): Follows from $W^{(t)}$ is column stochastic matrix.
\end{proof}

\subsection{Main Proof of Lemma 5}
\begin{proof}
To prove this lemma,we consider $t=0$ and $t\geq1$ separately. 

\noindent
\textbf{Case} $t=0$:
\begin{eqnarray}\label{lem_aux_aux_proof_1}
\overline{\textbf{z}}^{(t+1)}-\overline{\textbf{z}}^{(t)}&=&\overline{\textbf{z}}^{(1)}-\overline{\textbf{z}}^{(0)} \nonumber \\
~&\overset{Definition\ 3}{=}&\frac{1}{1-\beta} \overline{\textbf{x}}^{(1)}-\frac{\beta}{1-\beta} \overline{\textbf{x}}^{(0)}-\overline{\textbf{x}}^{(0)} \nonumber \\
~&=&\frac{1}{1-\beta} \left(\overline{\textbf{x}}^{(1)}-\overline{\textbf{x}}^{(0)}\right) \nonumber \\
~&\overset{Lemma\ 10}{=}&-\frac{\gamma}{1-\beta}\left(\beta \overline{\textbf{m}}^{(0)}+\frac{1}{N}\sum_{i=1}^{N}\textbf{g}_{i}^{(0)}\right) \nonumber \\
~&\overset{(a)}{=}& -\frac{\gamma}{1-\beta} \frac{1}{N} \sum_{i=1}^{N}\textbf{g}_{i}^{(0)}
\end{eqnarray}
(a): Follows from $\textbf{m}_{i}^{(0)}=\textbf{0}$ for $\forall i \in \{1,2,...,N\}$ in Algorithm 2.

\noindent
\textbf{Case} $t\geq1$:
\begin{eqnarray}\label{lem_aux_aux_proof_2}
\overline{\textbf{z}}^{(t+1)}-\overline{\textbf{z}}^{(t)}&\overset{Definition\ 3}{=}&\frac{1}{1-\beta}\left( \overline{\textbf{x}}^{(t+1)}-\overline{\textbf{x}}^{(t)}\right)-\frac{\beta}{1-\beta} \left( \overline{\textbf{x}}^{(t)}-\overline{\textbf{x}}^{(t-1)}\right) \nonumber \\
~&\overset{Lemma\ 10}{=}&\frac{1}{1-\beta} \left(-\gamma \overline{\textbf{m}}^{(t+1)}\right)-\frac{\beta}{1-\beta} \left(-\gamma \overline{\textbf{m}}^{(t)}\right) \nonumber \\
~&\overset{Lemma\ 10}{=}&-\frac{\gamma}{1-\beta} \left(\beta \overline{\textbf{m}}^{(t)}+\frac{1}{N}\sum_{i=1}^{N}\textbf{g}_{i}^{(t)}\right)+\frac{\beta\gamma}{1-\beta}  \overline{\textbf{m}}^{(t)} \nonumber \\
~&=&-\frac{\gamma}{1-\beta}\frac{1}{N}\sum_{i=1}^{N}\textbf{g}_{i}^{(t)}
\end{eqnarray}
\end{proof}

\section{Proof of Lemma 6}
\begin{proof}
Telescoping the first equation of Lemma 10 until $\overline{\textbf{m}}^{(0)}$ and considering $\overline{\textbf{m}}^{(0)}=\textbf{0}$, we have
\begin{equation}\label{lem_aux_x_proof_1}
\overline{\textbf{m}}^{(t)}=\sum_{s=0}^{t-1}\beta^{t-1-s}\left(\frac{1}{N}\sum_{i=1}^{N}\textbf{g}_{i}^{(s)}\right),\quad \forall t\geq 1
\end{equation}

\noindent
According to the definitions, for all $t\geq1$, we have
\begin{equation}\label{lem_aux_x_proof_2}
\overline{\textbf{z}}^{(t)}-\overline{\textbf{x}}^{(t)}=\frac{\beta}{1-\beta}\left(\overline{\textbf{x}}^{(t)}-\overline{\textbf{x}}^{(t-1)}\right)=-\frac{\beta\gamma}{1-\beta}\overline{\textbf{m}}^{(t)}
\end{equation}

\noindent
Combining \eqref{lem_aux_x_proof_1} and \eqref{lem_aux_x_proof_2} yields
\begin{equation}\label{lem_aux_x_proof_3}
\left \|\overline{\textbf{z}}^{(t)}-\overline{\textbf{x}}^{(t)}\right \|^{2}=\frac{\beta^{2}\gamma^{2}}{(1-\beta)^{2}}\left \|\sum_{s=0}^{t-1}\beta^{t-1-s}\left(\frac{1}{N}\sum_{i=1}^{N}\textbf{g}_{i}^{(s)}\right)\right \|^{2}
\end{equation}

\noindent
Define $h_{t}=\sum_{s=0}^{t-1}\beta^{t-1-s}=\frac{1-\beta^{t}}{1-\beta}$. For all $t\geq1$, we have
\begin{eqnarray}\label{lem_aux_x_proof_4}
\left \|\overline{\textbf{z}}^{(t)}-\overline{\textbf{x}}^{(t)}\right \|^{2}&=&\frac{\beta^{2}\gamma^{2}}{(1-\beta)^{2}}h_{t}^{2}\left \|\sum_{s=0}^{t-1}\frac{\beta^{t-1-s}}{h_{t}}\left(\frac{1}{N}\sum_{i=1}^{N}\textbf{g}_{i}^{(s)}\right)\right \|^{2} \nonumber \\
~&\overset{(a)}{\leq}&\frac{\beta^{2}\gamma^{2}}{(1-\beta)^{2}}h_{t}^{2}\sum_{s=0}^{t-1}\frac{\beta^{t-1-s}}{h_{t}}\left \|\frac{1}{N}\sum_{i=1}^{N}\textbf{g}_{i}^{(s)}\right \|^{2} \nonumber \\
~&\overset{(b)}{\leq}&\frac{\beta^{2}\gamma^{2}}{(1-\beta)^{3}}\sum_{s=0}^{t-1}\beta^{t-1-s}\left \|\frac{1}{N}\sum_{i=1}^{N}\textbf{g}_{i}^{(s)}\right \|^{2}
\end{eqnarray}
(a): The convexity of $\left \| \cdot \right \|^{2}$ and Jensen's inequality.\\
(b): Follows from $h_{t}=\frac{1-\beta^{t}}{1-\beta}\leq\frac{1}{1-\beta}$.

\noindent
Note that $\overline{\textbf{z}}^{(0)}-\overline{\textbf{x}}^{(0)}=\textbf{0}$, summing up (\ref{lem_aux_x_proof_4}) over $t\in \{1,2,...,T-1\}$, we have
\begin{eqnarray}\label{lem_aux_x_proof_5}
\sum_{t=0}^{T-1}\left \|\overline{\textbf{z}}^{(t)}-\overline{\textbf{x}}^{(t)}\right \|^{2}&\leq&\frac{\beta^{2}\gamma^{2}}{(1-\beta)^{3}}\sum_{t=1}^{T-1}\sum_{s=0}^{t-1}\beta^{t-1-s}\left \|\frac{1}{N}\sum_{i=1}^{N}\textbf{g}_{i}^{(s)}\right \|^{2}\nonumber \\
&=&\frac{\beta^{2}\gamma^{2}}{(1-\beta)^{3}}\sum_{s=0}^{T-2}\left(\left \|\frac{1}{N}\sum_{i=1}^{N}\textbf{g}_{i}^{(s)}\right \|^{2}\sum_{l=s+1}^{T-1}\beta^{l-1-s}\right)\nonumber \\
&\overset{(a)}{\leq}&\frac{\beta^{2}\gamma^{2}}{(1-\beta)^{4}}\sum_{s=0}^{T-2}\left \|\frac{1}{N}\sum_{i=1}^{N}\textbf{g}_{i}^{(s)}\right \|^{2} \nonumber \\
&\leq&\frac{\beta^{2}\gamma^{2}}{(1-\beta)^{4}}\sum_{s=0}^{T-1}\left \|\frac{1}{N}\sum_{i=1}^{N}\textbf{g}_{i}^{(s)}\right \|^{2}
\end{eqnarray}
(a): Follows by $\sum_{l=s+1}^{T-1}\beta^{l-1-s}=\frac{1-\beta^{T-s-1}}{1-\beta}\leq \frac{1}{1-\beta}$
\end{proof}

\section{Experiential Settings}
The settings of hyperparameters $v$ and $k$ of Section 6.B are illustrated as follows.

\begin{table}[htbp]\label{hyper}
\centering
\caption{$v$ and $k$}
\begin{tabular}{|cc|cccc|cccc|}
\hline
\multicolumn{2}{|c|}{Model + Dataset}            & \multicolumn{4}{c|}{Resnet18 + Cifar10}                                                     & \multicolumn{4}{c|}{Resnet50 + Cifar100}                                                       \\ \hline
\multicolumn{2}{|c|}{Topology}                   & \multicolumn{1}{c|}{Full} & \multicolumn{1}{c|}{Divide} & \multicolumn{1}{c|}{Exp} & Random & \multicolumn{1}{c|}{Full}  & \multicolumn{1}{c|}{Divide} & \multicolumn{1}{c|}{Exp}   & Random \\ \hline
\multicolumn{1}{|c|}{\multirow{2}{*}{SGAP}}  & k & \multicolumn{1}{c|}{0.1}  & \multicolumn{1}{c|}{0.01}   & \multicolumn{1}{c|}{1.0} & 0.1    & \multicolumn{1}{c|}{0.001} & \multicolumn{1}{c|}{0.001}  & \multicolumn{1}{c|}{0.001} & 0.001  \\ \cline{2-10} 
\multicolumn{1}{|c|}{}                       & v & \multicolumn{1}{c|}{0.1}  & \multicolumn{1}{c|}{0.1}    & \multicolumn{1}{c|}{0.1} & 0.1    & \multicolumn{1}{c|}{0.1}   & \multicolumn{1}{c|}{0.1}    & \multicolumn{1}{c|}{0.1}   & 0.1    \\ \hline
\multicolumn{1}{|c|}{\multirow{2}{*}{MSGAP}} & k & \multicolumn{1}{c|}{0.01} & \multicolumn{1}{c|}{0.01}   & \multicolumn{1}{c|}{0.1} & 0.01   & \multicolumn{1}{c|}{0.001} & \multicolumn{1}{c|}{0.001}  & \multicolumn{1}{c|}{0.001} & 0.001  \\ \cline{2-10} 
\multicolumn{1}{|c|}{}                       & v & \multicolumn{1}{c|}{0.1}  & \multicolumn{1}{c|}{0.1}    & \multicolumn{1}{c|}{0.1} & 0.1    & \multicolumn{1}{c|}{0.1}   & \multicolumn{1}{c|}{0.1}    & \multicolumn{1}{c|}{0.1}   & 0.1    \\ \hline
\end{tabular}
\end{table}

\bibliography{reference}

\begin{IEEEbiography}[{\includegraphics[width=1in,height=1.25in,clip,keepaspectratio]{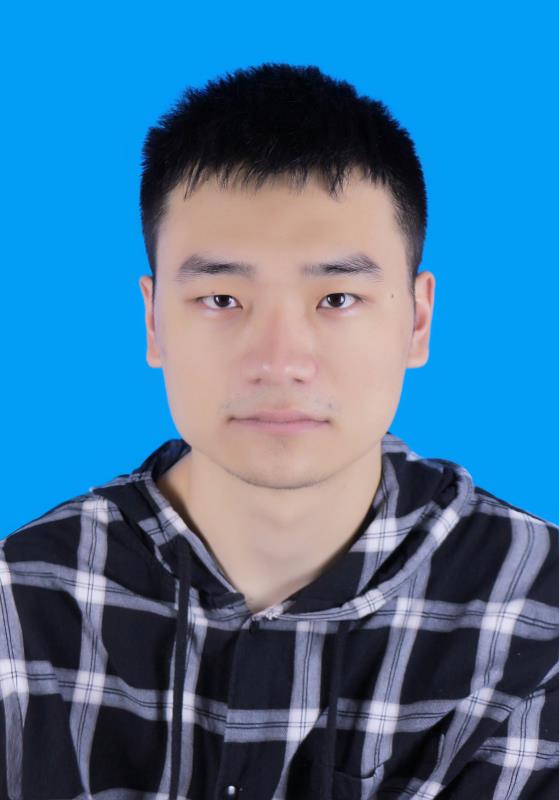}}]{Yiming Zhou}
received his bachelor's degree from the School of Mathematics, Taiyuan University of Technology in July, 2019. He is currently working toward the Ph.D degree from the School of Artificial Intelligence and Data Science, University of Science and Technology of China. His research interests include machine learning, distributed optimization, and network security.
\end{IEEEbiography}\vspace{-5em}

\begin{IEEEbiography}[{\includegraphics[width=1in,height=1.25in,clip,keepaspectratio]{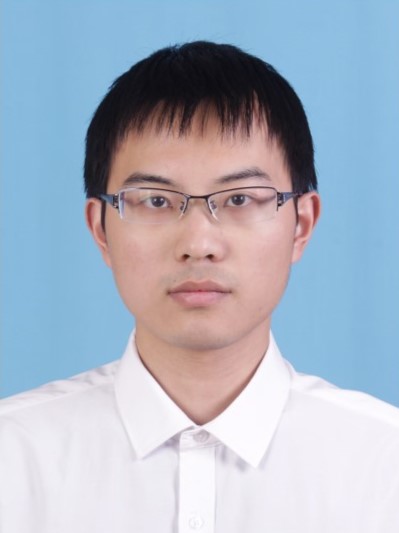}}]{Yifei Cheng}
received the B.S. degree from the University of Science and Technology of China (USTC), Hefei, China, in 2015. He is currently working towards the Ph.D. degree with USTC, Hefei, China. He is working with the Anhui Province Key Laboratory of Big Data Analysis and Application. His major research interests include optimization and federated learning.
\end{IEEEbiography}\vspace{-5em}

\begin{IEEEbiography}[{\includegraphics[width=1in,height=1.25in,clip,keepaspectratio]{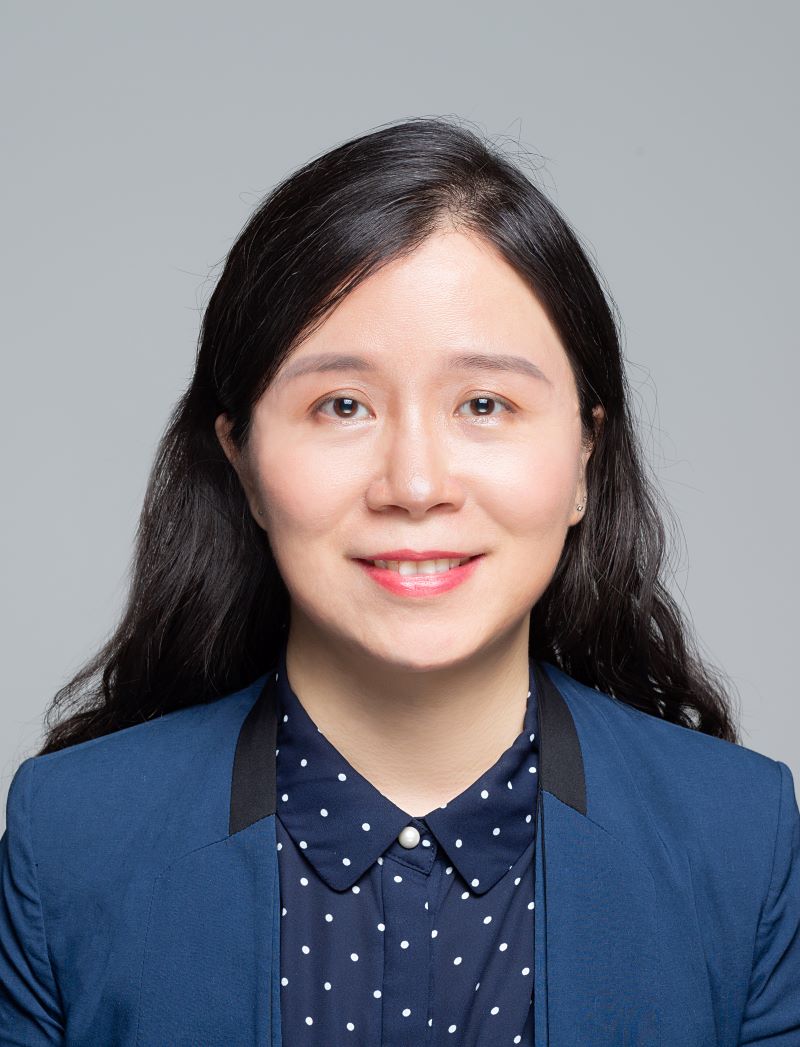}}]{Linli Xu}
is currently a professor of the School of Computer Science and Technology, USTC. Her research area is unsupervised learning and semi-supervised learning, multi-task learning and transfer learning, and optimization. She has published several papers in referred conference proceedings, such as AAAI, NeurIPS, and ICML.
\end{IEEEbiography}\vspace{-5em}

\begin{IEEEbiography}[{\includegraphics[width=1in,height=1.25in,clip,keepaspectratio]{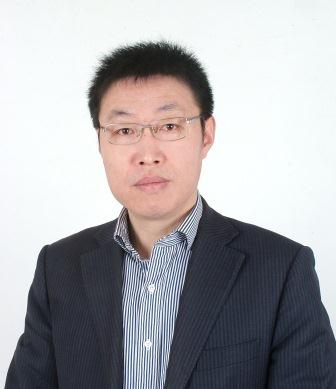}}]{Enhong Chen}
(Fellow, IEEE) received the B.S. degree from Anhui University in 1989, the M.S. degree from Hefei University of Technology in 1992, and the Ph.D. degree in Computer Science from the University of Science and Technology of China (USTC) in 1996. Currently, he is a Professor at the School of Computer Science and Technology, USTC. He is also a CCF Fellow, IEEE Fellow, and ACM Distinguished Member. He is the winner of the National Science Fund for Distinguished Young Scholars (2013) and a member of the Decision Advisory Committee of Shanghai (since June 2018). Additionally, he serves as the Vice Dean of the Faculty of Information and Intelligence at USTC, the Vice Director of the National Engineering Laboratory for Speech and Language Information Processing, the Director of the Anhui Province Key Laboratory of Big Data Analysis and Application, and the Chairman of the Anhui Province Big Data Industry Alliance. His current research interests include data mining and machine learning, particularly in social network analysis and recommendation systems. He has published more than 200 papers in various journals and conferences, including IEEE Transactions, ACM Transactions, and key data mining conferences such as KDD, ICDM, and NeurIPS. He has received several awards, including the Best Application Paper Award at KDD 2008, the Best Research Paper Award at ICDM 2011, the Outstanding Paper Award at FCS 2016, and the Best Student Paper Award at KDD 2024.
\end{IEEEbiography}

\end{document}